\documentclass[12pt]{article} 
\def\cdate{{February 8, 2017}}
\usepackage{amsfonts}
\usepackage{amsmath}
\usepackage{latexsym}
\usepackage{amssymb}
\usepackage{txfonts}
\usepackage[american]{babel}
\usepackage{tikz}
\usepackage{pgfplots}
\usepgfplotslibrary{groupplots}
\usetikzlibrary{intersections}

\makeatletter
\def\timenow{%
\@tempcnta=\time \divide\@tempcnta by 60 \number\@tempcnta:\multiply
\@tempcnta by 60 \@tempcntb=\time \advance\@tempcntb by -\@tempcnta
\ifnum\@tempcntb <10 0\number\@tempcntb\else\number\@tempcntb\fi}
\newcounter{outputpage}
\renewcommand{\@evenhead}{\raisebox{0pt}[\headheight][0pt]{\vbox{\hbox
to \textwidth{\thepage\hfil\strut\textit{\leftmark}}
}}}
\renewcommand{\@oddhead}{\raisebox{0pt}[\headheight][0pt]{\vbox{\hbox
to \textwidth{\textit{\rightmark}\hfil\strut\thepage}
}}}
\renewcommand{\@oddfoot}
{\vbox{
}}
\renewcommand{\@evenfoot}
{\vbox{
\vspace{3pt}
\hfil
{\scriptsize\textit{
\stepcounter{outputpage}
\hfill\hfill
\jobname.tex (\timenow, \today)
}}
\hfil
}}
\makeatother

\def\RR{{\mathbb R}}

\def\ZZ{{\mathbb Z}}

\def\d{\delta} 
\def\eps{\varepsilon} 
 
\def\m{\mu}

\def\d{\delta} 
\def\eps{\varepsilon}

\def\m{\mu}

\def\eps{\varepsilon}

\def\tr{\mathrm{tr}}

\def\tr{\mathrm{ tr\,}} 
\def\Tr{\mathrm{ Tr\,}}

\def\vol{\mathrm{ vol\,}}

\def\Spin{\mathrm{Spin}} 

\def\be{\begin{equation}} 
\def\ee{\end{equation}} 
\def\bea{\begin{eqnarray}} 
\def\eea{\end{eqnarray}} 
\def\bed{\begin{definition}{\ }}
\def\eed{\end{definition}}
\def\bd{\begin{description}}
\def\ed{\end{description}}
\def\bc{\begin{center}}
\def\ec{\end{center}}

\newtheorem{lemma}{Lemma}
\newtheorem{corollary}{Corollary}

\newtheorem{definition}{Definition}
\def\sideremark#1{\ifvmode\leavevmode\fi\vadjust{\vbox to 0pt{\vss\hbox to 0pt
{\hskip\hsize\hskip1em\vbox{\hsize2cm\tiny\raggedright\pretolerance10000
\noindent #1\hfill}\hss}\vbox to8pt{\vfil}\vss}}}

\begin{document}


\begin{titlepage}
\thispagestyle{empty}
\null
\vspace{-3cm}
\hspace*{50truemm}{\hrulefill}\par\vskip-4truemm\par
\hspace*{50truemm}{\hrulefill}\par\vskip5mm\par
\hspace*{50truemm}{{\large\sc New Mexico Tech {\rm
(\cdate)
}}}
\par
\hspace*{50truemm}{\hrulefill}
\par\vskip-4truemm\par
\hspace*{50truemm}{\hrulefill}
\par
\bigskip
\bigskip
\par
\par
\vspace{3cm}
\null
\centerline{\huge\bf One-Loop Quantum Gravity}
\bigskip
\centerline{\huge\bf in the Einstein Universe}
\bigskip
\bigskip
\bigskip
\centerline{\Large\bf Ivan G. Avramidi and Samuel J. Collopy}
\bigskip
\centerline{\it Department of Mathematics}
\centerline{\it New Mexico Institute of Mining and Technology}
\centerline{\it Socorro, NM 87801, USA}
\centerline{\it E-mail: iavramid@nmt.edu, samuel.collopy@gmail.com}
\bigskip
\bigskip
\centerline{\cdate}
\vfill
{\narrower
\par

We study quantum gravity with the
Einstein-Hilbert action including the cosmological constant
on the Euclidean Einstein universe $S^1\times S^3$. 
We compute exactly the spectra and the heat kernels of the relevant operators on $S^3$
and  use these results to compute the heat trace of the graviton and ghost
operators and the exact one-loop effective action on $S^1\times S^3$. 
We show that the system is unstable in the infrared limit
due to the presence of the negative modes of the graviton and the ghost operators.  
We study the thermal properties of the model with the temperature
$T=(2\pi a_1)^{-1}$
determined by the radius $a_1$ of the circle $S^1$.
We show that the heat capacity $C_v$ is well defined and behaves like $\sim T^3$
in the high temperature limit and has a singularity of the type 
$\sim (T-T_c)^{-1}$, indicating a second-order phase transition,
with the critical temperature $T_c$ 
determined by the cosmological constant $\Lambda$ and the radius $a$ of the 
sphere $S^3$. We also discuss some peculiar properties of the model such as 
the negative heat capacity as well as possible physical applications. 
}


\end{titlepage}

\section{Introduction}
\setcounter{equation}{0}

The low-energy effective action in quantum field theory is a powerful
tool that enables one to study the vacuum state of the theory. 
The low-energy effective action  cannot be computed
in the usual perturbation theory, 
and so to study it in
the generic case, one needs new essentially non-perturbative methods. The
development of such methods for the calculation of the heat kernel was
initiated in our papers 
\cite{avramidi93,avramidi95a} for a gauge theory in
flat space, which were then applied to study the vacuum structure of the
Yang-Mills theory in 
\cite{avramidi95b,avramidi99}. 
These ideas were first
extended to scalar fields on curved manifolds in 
\cite{avramidi94,avramidi96}
and finally to arbitrary twisted spin-tensor fields in
\cite{avramidi09a}.
In \cite{avramidi10} we applied these methods to study quantum gravity
and Yang-Mills theory on any symmetric space.
Further, we applied these methods to study the thermal Yang-Mills theory on product of spheres,
such as $S^1\times S^1\times S^2$ and $S^1\times S^3$ in \cite{avramidi12a, avramidi12b}.

In the present paper we apply these methods to study the one-loop low-energy
effective action in quantum Einstein general relativity in the Einstein Universe background
at finite temperature.
{}From the mathematical point of view, we compute the one-loop effective action
for the Einsten-Hilbert action with cosmological constant on the background
$S^1\times S^3$.

This paper is organized as follows. In Sec. 2 we introduce all the relevant operators
for the calculation of the one-loop effective action in Einstein quantum gravity.
We refer to the paper \cite{avramidi10}
for the details. In Sec. 3 we compute the 
heat trace coefficients.
In Sec 4. we study the quantum gravity on $S^3$ and compute
all relevant heat traces on $S^3$.
We refer to the paper \cite{avramidi12b} for the details of the
calculation of the heat traces on $S^3$ for any representation. 
In Sec. 5 we compute the heat traces and the effective action on $S^1\times S^3$.
Finally, in Sec. 6 we discuss the thermodynamic properties of the model.

\section{One-loop Einstein Gravity}
\setcounter{equation}{0}

In this section we follow our previous work \cite{avramidi10,avramidi10b}.
The dynamics of the gravitational field parametrized by the Riemannian metric
on a closed (compact without boundary) manifold $(M,g)$
of dimension $n$
is described by the Hilbert-Einstein action of general relativity,
which (in Euclidean formulation) has the form
\be
S=\frac{1}{16\pi G}\int\limits_M dx\;g^{1/2}
\left(-R+2\Lambda\right)\;,
\ee
where $g=\det g_{\mu\nu}$,
$G$ is the gravitational constant and $\Lambda$ is the cosmological
constant. 
The classical vacuum Einstein equations are determined by the first
variation of the action
\be
16\pi G g^{-1/2}\frac{\delta S}{\delta g_{\mu\nu}}
=R^{\mu\nu}-\frac{1}{2}Rg^{\mu\nu}+\Lambda g^{\mu\nu}=0\,.
\ee

In two dimensions the action is trivial
\be
S=\frac{1}{16\pi G}\left\{
-4\pi \chi(M)+2\Lambda \vol(M)\right\},
\ee
where $\chi(M)$ is the Euler characteristic of the manifold $M$
and $\vol(M)$ is its volume. Therefore, it does not have any extremal
metrics; more precisely, in two dimensions every metric satisfies the 
Einstein equations with zero cosmological constant,
\be
R_{ab}=\frac{1}{2}Rg_{ab},
\ee
and, therefore, the Einstein equations do not have any solutions
for any $\Lambda\ne 0$, which means that Einstein gravity
in two dimensions is purely topological.

For this reason, we restrict ourselves to $n>2$.
In this case the Riemann tensor can be decomposed as follows
\bea
R^{ab}{}_{cd}&=&C^{ab}{}_{cd}
+\frac{4}{n-2} R^{[a}{}_{[c}\delta^{b]}{}_{d]}
-\frac{2}{(n-1)(n-2)} R \delta^{[a}{}_{[c}\delta^{b]}{}_{d]},
\eea
where $C_{abcd}$ is the Weyl tensor.
The norm of the Riemann tensor is then
\bea
R_{abcd}R^{abcd}=C_{abcd}C^{abcd}
+\frac{4}{n-2} R_{ab}R^{ab}-\frac{2}{(n-1)(n-2)} R^2.
\eea
The solutions of the Einstein equations
determine the Einstein
spaces, 
\be
R_{ab}=\frac{2}{n-2}\Lambda g_{ab}\,,
\label{mass-shell}
\ee
and, therefore,
\be
R=\frac{2n}{n-2}\Lambda\,.
\label{mass-shell1}
\ee
In this case the Riemann tensor is
\bea
R^{ab}{}_{cd}&=&C^{ab}{}_{cd}
+\frac{4}{(n-1)(n-2)} \Lambda \delta^{[a}{}_{[c}\delta^{b]}{}_{d]},
\eea
with the norm
\bea
R_{abcd}R^{abcd}=C_{abcd}C^{abcd}
+\frac{8n}{(n-1)(n-2)^2}\Lambda^2.
\eea

The case of three dimensions is special. In this case 
the Weyl tensor is equal to zero identically, and,
therefore, the Riemann tensor is 
fully determined by the Ricci tensor, 
\bea
R^{ab}{}_{cd} &=& 
4R^{[a}{}_{[c}\delta^{b]}{}_{d]}
-R\delta^{[a}{}_{[c}\delta^{b]}{}_{d]}.
\label{27aa}
\eea
Therefore, in particular,
\be
R_{abcd}R^{abcd}- 4R_{ab}R^{ab}+R^2=0\,.
\label{2.8xx}
\ee
The Einstein equations take the form
\be
R_{ab} = 2\Lambda g_{ab}
\ee
and, therefore, the curvature tensor of Einstein spaces is 
fully determined by the metric, 
\bea
R^{ab}{}_{cd} &=& 
2\Lambda \delta^{[a}{}_{[c}\delta^{b]}{}_{d]}
\,.
\eea
This means that the only Einstein spaces in three dimensions
are the (locally)
maximally symmetric spaces,
the sphere $S^3$ for $\Lambda>0$, the hyperbolic manifolds
$H^3/\Gamma$ for $\Lambda<0$, where $\Gamma$ is a lattice in $SO^+(1,3)$;
for $\Lambda=0$ the only solutions are flat manifolds, 
like a torus $T^3$.
In any case, gravity in three dimensions is rigid, that is, it
does not have any propagating
degrees of freedom.

Notice that the same invariant (\ref{2.8xx}) plays a role in higher dimensions as well.
In particular, in dimension $n=4$ the integral of that invariant 
determines the Euler 
characteristic of the manifold
\bea
\chi(M) &=&
\frac{1}{32\pi^2}\int\limits_M dx\;g^{1/2} \left(R_{abcd}R^{abcd}
-4R_{ab}R^{ab}+R^2\right)
\nonumber\\
&=&
\frac{1}{32\pi^2}\int\limits_M dx\;g^{1/2} 
\left(C_{abcd}C^{abcd}
-2R_{ab}R^{ab}+\frac{2}{3}R^2\right).
\eea
When the Einstein equations (\ref{mass-shell}),(\ref{mass-shell1}), are satisfied 
the Ricci tensor is determined by the metric. That is, for $n=4$,
\be
R_{ab}=\Lambda g_{ab}, \qquad R=4\Lambda,
\ee
and the integral norm of the Riemann tensor is determined by
the Euler characteristic
\bea
\chi(M) &=&\frac{1}{32\pi^2}
\int\limits_M dx\;g^{1/2}R_{abcd}R^{abcd}
\nonumber\\
&=&
\frac{1}{32\pi^2}
\int\limits_M dx\;g^{1/2}\left(C_{abcd}C^{abcd}
+\frac{8}{3}\Lambda^2\right).
\eea
Note that the Euler characteristic of Einstein spaces 
in four dimensions is positive definite.
It is worth stressing that this disagrees with eq. (115)
in \cite{hawking79}.

The diffeomorphism invariance of the Einstein-Hilbert functional means
that the metric carries some non-physical (gauge) degrees of freedom
described by a vector field. 
In $n$ dimensions 
a vector field has $n$ independent components
and a symmetric 2-tensor field 
has
$n(n+1)/2$ independent components. Therefore, the gravitational field
in $n$ dimensions has 
\be
N(n)=\frac{n(n+1)}{2}-2n=\frac{n(n-3)}{2}
\ee
degrees of freedom. This number is equal to $N(4)=2$ in four dimensions
as expected; however, it vanishes in three dimensions, $N(3)=0$. 
 In two dimensions it gives a meaningless result, $N(2)=-1$.
We will compute the effective action in three dimensions below
but one should realize that
in three dimensions the Einstein gravity does not have any dynamics
\cite{witten88,witten07}.

One of the fundamental problems of quantum Einstein gravity is that
the Euclidean Einstein-Hilbert action is unbounded from below, which leads
to the divergence of the Euclidean path integral over all metrics.
This divergence is conceptual in nature and is much more serious than the
usual ultraviolet divergence of the quantum field theory.
It is well
known \cite{hawking79}
that under a conformal transformation
\be
\bar g_{\mu\nu}=\omega^{4/(n-2)} g_{\mu\nu},
\ee
where $\omega$ is a smooth positive function on $M$,
the action takes the form
\be
S=\frac{1}{16\pi G}\frac{8(n-1)}{(n-2)}\int\limits_M dx\;g^{1/2}\left\{
-\frac{1}{2}\omega Y\omega
+\frac{(n-2)}{4(n-1)}\Lambda \omega^{2n/(n-2)}
\right\},
\ee
where $Y$ is the Yamabe operator
\be
Y=-\Delta+\frac{n-2}{4(n-1)}R.
\ee
The Yamabe operator is nothing but the conformally covariant scalar Laplacian.
It is a self-adjoint elliptic partial differential 
operator with a positive leading symbol. The spectrum of such operator
is real, discrete, and with finite multiplicities; it is 
bounded from below and unbounded from above.
This shows that the action functional is unbounded from below.
It is obvious that by keeping the metric $g_{\mu\nu}$ constant and taking
the function $\omega$ to be bounded and increasingly oscillating
the action can be made arbitrarily large and negative.
This is a well known conformal problem of quantum gravity.
It has been suggested \cite{hawking79}
that this problem can be avoided by deforming the contour
of integration in the path integral over the conformal factor
to make it purely imaginary, which will turn the action into
a standard functional of quantum field theory. However, such
an approach cannot be taken seriously. This is a major problem
of Einstein quantum gravity and it remains open. A solution to this
problem would require a modification of the Einstein-Hilbert action
but we do not attempt to solve it in the present paper.

The standard loop expansion of the Euclidean effective action
has the form
\be
\Gamma=S+\hbar\Gamma_{(1)}+O(\hbar^2),
\ee
where $\Gamma_{(1)}$ is the one-loop effective action.
The one-loop effective action is determined by
the graviton operator $L_2$ acting on symmetric two-tensor fields
and the Faddeev-Popov ghost operator  $L_1$ acting on vector fields.
In the Euclidean formulation the zeta-regularized
one-loop effective action has the form
\be
\Gamma_{(1)}=-\frac{1}{2}\zeta'_{GR}(0)\,,
\ee
where 
\be
\zeta_{GR}(s)=\zeta_{L_2}(s)-2\zeta_{L_1}(s)\,,
\ee
and
$\zeta_{L_1}(s)$ and $\zeta_{L_2}(s)$ are the zeta
functions of the operators $L_1$ and $L_2$ defined by
\be
\zeta_{L}(s)=\frac{\mu^{2s}}{\Gamma(s)}
\int\limits_0^\infty dt\; t^{s-1}e^{-tz^2}
\Theta_L(t)\,,
\ee
where 
\be
\Theta_L(t)=\Tr\exp(-tL).
\ee
The renormalization parameter $\mu$ is introduced to preserve dimensions
and $z$ is a sufficiently large {\it infra-red regularization parameter},
which should be set to zero at the end of the calculation.
Therefore,
\be
\zeta_{GR}(s)=\frac{\mu^{2s}}{\Gamma(s)}
\int\limits_0^\infty dt\; t^{s-1}e^{-tz^2}
\Theta_{GR}(t)\,,
\ee
where 
\be
\Theta_{GR}(t)=
\Theta_{L_2}(t)-2\Theta_{L_1}(t)\,;
\label{25xx}
\ee
we will call this invariant the heat trace of quantum gravity.

The operators $L_2$ and $L_1$ are determined by the second
variation of the action and then by imposing some gauge
condition on the metric fluctuation (see, for example, 
\cite{dewitt03,avramidi00}).
The second variation of the action defines a second-order
partial differential operator $P$ acting on symmetric
two-tensors by
\be
16\pi G g^{-1/2}\frac{\delta^2 S}{\delta g_{\mu\nu}\delta g_{\alpha\beta}}
h_{\alpha\beta}
=\frac{1}{2}P^{\mu\nu\alpha\beta}h_{\alpha\beta}\,,
\ee
where
\bea
P^{\mu\nu,\alpha\beta}
&=&
-\left(g^{\alpha(\mu}g^{\nu)\beta}
-g^{\alpha\beta}g^{\mu\nu}\right)\Delta
\nonumber\\[10pt]
&&
-g^{\mu\nu}\nabla^{(\alpha}\nabla^{\beta)}
-g^{\alpha\beta}\nabla^{(\mu}\nabla^{\nu)}
+2\nabla^{(\mu}g^{\nu)(\alpha}\nabla^{\beta)}
\nonumber\\[10pt]
&&
-2R^{(\mu|\alpha|\nu)\beta}
-g^{\alpha(\mu}R^{\nu)\beta}
-g^{\beta(\mu}R^{\nu)\alpha}
+R^{\mu\nu}g^{\alpha\beta}
+ R^{\alpha\beta}g^{\mu\nu}
\nonumber\\
&&
+\left(g^{\mu(\alpha}g^{\beta)\nu}
-\frac{1}{2}g^{\mu\nu}g^{\alpha\beta}\right)
(R-2\Lambda).
\eea

In the minimal gauge 
the non-diagonal derivatives in both the graviton
operator and the ghost operator vanish and the operators
take the form
\bea
\tilde L_2{}^{cd,ab}
&=&
\left(g^{a(c}g^{d)b}
-\frac{1}{2}
g^{ab}g^{cd}\right)(
-\Delta+R-2\Lambda)
\nonumber\\
&&
-2R^{(c|a|d)b}
-g^{a(c}R^{d)b}
-g^{b(c}R^{d)a)}
+R^{cd}g^{ab}
+g^{cd}R^{ab}
\,,
\\[10pt]
\tilde L_1{}^{ab}&=&-g^{ab}\Delta-R^{ab}.
\eea
We should stress that the operator $\tilde L_2$ differs from the 
eq. (16.37) in \cite{dewitt84}.

The tensor 
\be
E^{cd,ab}=g^{a(c}g^{d)b}
-\frac{1}{2}g^{ab}g^{cd}
\ee
here is the metric
in the space of symmetric tensors.
It is easy to see that it is positive definite in the subspace
of traceless symmetric tensors but it is
negative definite in the conformal (scalar) sector.
This is exactly the problem of the conformal mode in quantum gravity
discussed above. Following the standard approach 
\cite{hawking79,christensen80,dewitt84,dewitt03}
we simply assume
that it can be fixed somehow by some physical arguments and
proceed as follows.
We factor out this metric from the operator $\tilde L_2$
to define the graviton operator
$L_2$ and the ghost operator $L_1$
in the canonical Laplace-type form
\bea
L_j&=&-\Delta+Q_j\;,
\label{2.34xx}
\eea
where the potentials for both operators are
\cite{avramidi10}
\bea
\left(Q_{1}\right)^a{}_b
&=&-R^a{}_b\,,
\label{2.30xx}
\\[5pt]
\left(Q_{2}\right)^{cd}{}_{ab}
&=&
-2R^c{}_{(a}{}^d{}_{b)}
-2\delta^{(c}{}_{(a}R^{d)}{}_{b)}
+R^{cd}g_{ab}
+\frac{2}{n-2}g^{cd}R_{ab}
\nonumber\\
&&
-\frac{1}{(n-2)}g^{cd}g_{ab}R
+\delta^c{}_{(a}\delta^d{}_{b)}(R-2\Lambda)
\,.
\label{2.31xx}
\eea
We should stress here that the endomorphism $Q_2$ does not coincide with the 
eq. (16.78) in \cite{dewitt84}.

It is well known that the heat trace of Laplace type operators
has the asymptotic expansion as $t\to 0$
\be
\Theta_L(t)\sim(4\pi t)^{-n/2}\sum_{k=0}^\infty
t^k B_k(L),
\ee
where $B_k(L)$ are the so-called 
Hadamard-Minakshisundaram-DeWitt-Seeley coefficients
(or simply heat trace coefficients) of the operator $L$.
This means that the function $\Theta(t)$ has similar asymptotic
expansion as $t\to 0$
\be
\Theta_{GR}(t)\sim(4\pi t)^{-n/2}\sum_{k=0}^\infty
t^k C_k,
\label{2.38xx}
\ee
where
\be
C_k=B_k(L_2)-2B_k(L_1)\,.
\label{2.35xx}
\ee

It is easy to find the dependence of the effective action on the
renormalization parameter; by integrating the equation
\be
\mu\frac{\partial}{\partial\mu}\Gamma_{(1)}=-\zeta_{GR}(0)\,,
\ee
we get
\be
\Gamma_{(1)}(\mu)=\Gamma_{(1)}(\mu_0)-\log\left(\frac{\mu}{\mu_0}\right)\zeta_{GR}(0)\,.
\ee
This enables one to study the high-energy
asymptotics of the effective action as $\mu\to \infty$.

For the Laplace type operators the value of the zeta function
at $s=0$ is determined by a specific heat trace coefficient
\bea
\zeta_{L}(0)
=\left\{
\begin{array}{ll}
(4\pi)^{-n/2}B_{n/2}(L),  & \mbox{for even} \ n,
\\[5pt]
0, & \mbox{for odd} \ n.
\end{array}
\right.
\eea
Therefore,
\bea
\zeta_{GR}(0) 
=\left\{
\begin{array}{ll}
(4\pi)^{-n/2}C_{n/2}, & \mbox{for even} \ n,
\\[5pt]
0, & \mbox{for odd} \ n.
\end{array}
\right.,
\eea
in particular, in four dimensions, $n=4$,
\bea
\zeta_{GR}(0)=
(4\pi)^{-2}C_{2}.
\eea

\section{Heat Trace Coefficients}
\setcounter{equation}{0}

We will need the heat trace coefficients $B_0, B_1$ and $B_2$
for the operators $L_1$ and $L_2$.
They have the following well-known form
\cite{gilkey75,avramidi00} (we neglected the inessential total derivatives here
which do not contribute to the global invariants)
\bea
B_0(L) &=& \int\limits_M dx\;g^{1/2}\tr I,
\\
B_1(L) &=& \int\limits_M dx\;g^{1/2}\tr\left(\frac{1}{6}RI-Q\right),
\\
B_2(L) &=& \int\limits_M dx\;g^{1/2}\tr\Biggl\{
\frac{1}{2}Q^2-\frac{1}{6}RQ
+\frac{1}{12}{\cal R}_{ab}{\cal R}^{ab}
\nonumber\\
&&
+I\left(\frac{1}{72}R^2
+\frac{1}{180}R_{abcd}R^{abcd}
-\frac{1}{180}R_{ab}R^{ab}\right)
\Biggr\}.
\eea
Here $I$ is the identity endomorphism and ${\cal R}_{ab}$
is the curvature of the spin connection of a tensor field realizing 
a representation of the spin group
defined by
\be
{\cal R}_{\mu\nu}=\frac{1}{2}R^{ab}{}_{\mu\nu}\Sigma_{ab},
\ee
where $\Sigma_{ab}$ are the generators of the spin group $\Spin(n)$
satisfying the commutation relations
\be
[\Sigma_{ab},\Sigma_{cd}]=-g_{ac}\Sigma_{bd}
+g_{bc}\Sigma_{ad}
+g_{ad}\Sigma_{bc}
-g_{bd}\Sigma_{ac}.
\label{3.5xx}
\ee

For the vector representation the identity and the generators have the form
\bea
(I_{1})^c{}_d &=& \delta^c{}_d,
\\
(\Sigma_{(1), ab})^c{}_d &=& 2 \delta^c{}_{[a}g_{b]d},
\label{3.7xx}
\eea
Therefore, $\tr I_1=n$ and 
\bea
B_0(L_1) &=& n\vol(M).
\eea
Also, we have
\bea
\tr Q_1 &=& -R,
\eea
and, therefore,
\bea
B_1(L_1) &=& \int\limits_M dx\;g^{1/2} \frac{1}{6}(n+6)R.
\eea

Further, we compute
\bea
\tr \Sigma_{(1), ab} \Sigma_{(1)}{}^{pq} &=& 
-4\delta^{[p}{}_{[a}\delta^{q]}{}_{b]}
\eea
and
\bea
\tr {\cal R}_{(1), ab}{\cal R}_{(1)}{}^{ab} &=& - R_{abcd}R^{abcd}.
\eea
We also have
\bea
\tr (Q_1)^2 &=& R_{ab}R^{ab},
\eea
and, therefore,
\be
B_2(L_1) = \int\limits_M dx\;g^{1/2} 
\left\{
\frac{n-15}{180}R_{abcd}R^{abcd}
+\frac{90-n}{180}R_{ab}R^{ab}
+\frac{n+12}{72} R^2
\right\}.
\ee

Next, we compute the heat trace coefficients for the operator $L_2$.
For the tensor representation the identity and the generators have the form

\bea
(I_2)^{ef}{}_{cd}&=& \delta^{(e}{}_{(c}\delta^{f)}{}_{d)},
\\
(\Sigma_{(2), ab})^{ef}{}_{cd} &=& 
4\delta^{(e}{}_{[a}g_{b](c}\delta^{f)}{}_{d)}.
\label{3.16xx}
\eea
First, we have $\tr I_2=n(n+1)/2$, hence,
\bea
B_0(L_2) &=& \frac{1}{2}n(n+1)\vol(M).
\eea

Now, we introduce the following endomorphisms
\bea
(V_{1})^{cd}{}_{ab}&=&R^c{}_{(a}{}^d{}_{b)},
\\
(V_{2})^{cd}{}_{ab}&=&\delta^{(c}{}_{(a}R^{d)}{}_{b)},
\\
(V_{3})^{cd}{}_{ab}&=&R^{cd}g_{ab},
\\
(V_{4})^{cd}{}_{ab}&=&g^{cd}R_{ab},
\\
(V_{5})^{cd}{}_{ab}&=&g^{cd}g_{ab}.
\eea
Then the endomorphism $Q_2$ has the form
\be
Q_2=\tilde Q_2+(R-2\Lambda)I_2,
\ee
where
\be
\tilde Q_2=-2V_1-2V_2+V_3
+\frac{2}{n-2}V_4
-\frac{1}{(n-2)}RV_5.
\ee

Now, by using the traces
\bea
\tr V_1 &=& -\frac{1}{2}R,
\\
\tr V_2 &=&\frac{1}{2}(n+1)R,
\\
\tr V_3 &=&R,
\\
\tr V_4&=&R,
\\
\tr V_5&=&n.
\eea
we compute
\be
\tr \tilde Q_2=-nR,
\ee
and, therefore,
\bea
\tr Q_2 &=&\frac{1}{2}n(n-1)R-n(n+1)\Lambda,
\eea
which gives
\bea
B_1(L_2) &=& \int\limits_M dx\;g^{1/2}\left\{
-\frac{1}{12}n(5n-7)R
+n(n+1)\Lambda
\right\}.
\eea

Now, we have
\bea
\tr (Q_2)^2 
&=&\tr(\tilde Q_2)^2
+\frac{1}{2}n(n-3)R^2
-2n(n-1)R\Lambda
+2n(n+1)\Lambda^2.
\eea
Next, by using the traces of the products
\bea
\tr V_1V_1&=&\frac{3}{4}R_{abcd}R^{abcd},
\\
\tr V_2V_2&=&\frac{1}{4}(n+2)R_{ab}R^{ab}+\frac{1}{4}R^2,
\\
\tr V_3V_3&=&R^2,
\\
\tr V_4V_4&=&R^2,
\\
\tr V_5V_5&=&n^2,
\\
\tr V_1V_2&=&-\frac{1}{2}R_{ab}R^{ab},
\\
\tr V_1V_3&=&R_{ab}R^{ab},
\\
\tr V_1V_4&=&R_{ab}R^{ab},
\\
\tr V_1V_5&=&R,
\\
\tr V_2V_3&=&R_{ab}R^{ab},
\\
\tr V_2V_4&=&R_{ab}R^{ab},
\\
\tr V_2V_5&=&R,
\\
\tr V_3V_4&=&nR_{ab}R^{ab},
\\
\tr V_3V_5&=&nR,
\\
\tr V_4V_5&=&nR,
\eea
we obtain
\bea
\tr (\tilde Q_2)^2 &=&
3R_{abcd}R^{abcd}
+\frac{n^2-8n+4}{n-2}R_{ab}R^{ab}
+\frac{n+2}{n-2}R^2.
\eea
By using these results we get
\bea
\tr (Q_2)^2 &=&
3R_{abcd}R^{abcd}
+\frac{n^2-8n+4}{n-2}R_{ab}R^{ab}
+\frac{n^3-5n^2+8n+4}{2(n-2)}R^2
\nonumber\\
&&
-2n(n-1)R\Lambda
+2n(n+1)\Lambda^2.
\eea

We introduce yet another endomorphism
\bea
(T_{pq})^{cd}{}_{ab} &=& \delta^{(c}{}_{[p}g_{q](a}\delta^{d)}{}_{b)}.
\eea
By using
\bea
\tr T_{pq}T^{rs}&=&-\frac{1}{4}(n+2)\delta^{[r}{}_{[p}\delta^{s]}{}_{q]}
\eea
we get 
\bea
\tr \Sigma_{(2), ab} \Sigma_{(2)}{}^{pq} &=&
-4(n+2)\delta^{[p}{}_{[a}\delta^{q]}{}_{b]},
\eea
and, therefore,
\bea
\tr {\cal R}_{(2), ab}{\cal R}_{(2)}{}^{ab} &=&
-(n+2)R_{abcd}R^{abcd}.
\eea

By using these results we get
\be
B_2(L_2) = \int\limits_M dx\;g^{1/2}
\Biggl\{
\alpha_1R_{abcd}R^{abcd}
+\alpha_2R_{ab}R^{ab}
+\alpha_3R^2
+\gamma_1R\Lambda
+\gamma_2\Lambda^2
\Biggr\},
\ee
where
\bea
\alpha_1&=&\frac{1}{360}(n^2-29n+480),
\\[5pt]
\alpha_2&=&\frac{-n^3+181n^2-1438n+720}{360(n-2)},
\\[5pt]
\alpha_3&=&\frac{25n^3-145n^2+262n+144}{144(n-2)},
\\[5pt]
\gamma_1&=&\frac{1}{6}n(7-5n),
\\[5pt]
\gamma_2&=&n(n+1).
\eea

Finally, by using these results we compute the coefficients $C_k$,
(\ref{2.35xx}),
\bea
C_0 &=&\frac{1}{2}n(n-3)\vol(M),
\label{3.61xx}
\\
C_1 &=& \int\limits_M dx\;g^{1/2}\left\{
-\frac{1}{12}(5n^2-3n+24)R
+n(n+1)\Lambda
\right\},
\\
C_2 &=& \int\limits_M dx\;g^{1/2}
\left\{
\beta_1R_{abcd}R^{abcd}
+\beta_2R_{ab}R^{ab}
+\beta_3R^2
+\gamma_1R\Lambda
+\gamma_2\Lambda^2
\right\},
\nonumber\\
\label{3.63xx}
\eea
where
\bea
\beta_1&=&\frac{1}{360}(n^2-33n+540),
\\[5pt]
\beta_2&=&\frac{-n^3+185n^2-1806n+1440}{360(n-2)},
\\[5pt]
\beta_3&=&\frac{25n^3-149n^2+222n+240}{144(n-2)}.
\eea
These results are also different from the ones given by eqs. (16.79)-(16.81)
of \cite{dewitt84}.


We will  need these coefficients for $n=4$. 
By using the above results we obtain
in this case
\bea
B_0(L_1) &=& 4\vol(M),
\\[5pt]
B_1(L_1) &=& \int\limits_M dx\;g^{1/2} \frac{5}{3}R,
\\[5pt]
B_2(L_1) &=& \int\limits_M dx\;g^{1/2} 
\left\{
-\frac{11}{180}R_{abcd}R^{abcd}
+\frac{43}{90}R_{ab}R^{ab}
+\frac{2}{9} R^2
\right\},
\eea
\bea
B_0 (L_2)&=& 10\vol(M),
\\[5pt]
B_1(L_2) &=& \int\limits_M dx\;g^{1/2}\left\{
-\frac{13}{3}R
+20\Lambda
\right\},
\\[5pt]
B_2(L_2) &=& \int\limits_M dx\;g^{1/2}
\Biggl\{
\frac{19}{18}R_{abcd}R^{abcd}
-\frac{55}{18}R_{ab}R^{ab}
+\frac{59}{36}R^2
-\frac{26}{3}R\Lambda
+20\Lambda^2
\Biggr\}.
\nonumber\\
\eea
Therefore, we obtain the total coefficients (\ref{2.35xx})
\bea
C_0 &=&2\vol(M),
\\[5pt]
C_1 &=& \int\limits_M dx\;g^{1/2}\left\{
-\frac{23}{3}R
+20\Lambda
\right\},
\\[5pt]
C_2 &=& \int\limits_M dx\;g^{1/2}
\left\{
\frac{53}{45}R_{abcd}R^{abcd}
-\frac{361}{90}R_{ab}R^{ab}
+\frac{43}{36}R^2
-\frac{26}{3}R\Lambda
+20\Lambda^2
\right\}
\nonumber\\
&=& 
\frac{1696}{45}\pi^2\chi(M)+
\int\limits_M dx\;g^{1/2}
\left\{-\frac{1}{90}R_{ab}R^{ab}
+\frac{7}{36}R^2
-\frac{26}{3}R\Lambda
+20\Lambda^2
\right\}\,.
\nonumber\\
\eea
Finally, when the Einstein equations are satisfied these coefficients 
in four dimensions take the form
\bea
C_0 &=& 2\vol(M),
\\[5pt]
C_1 &=& -\frac{32}{3}\Lambda\vol(M),
\\[5pt]
C_2
&=& \frac{1696}{45}\pi^2\chi(M)
-\frac{58}{5}\Lambda^2\vol(M).
\eea
This gives the quantity 
\be
\zeta_{GR}(0)=\frac{106}{45}\chi(M)
-\frac{29}{40}\frac{\Lambda^2}{\pi^2}\vol(M),
\ee
determining the scaling properties of the model.
It is worth stressing that this result 
coincides with the eq. (4.23) of
\cite{christensen80} and disagrees with the results
of \cite{hawking79}, eq. (79), and \cite{gibbons78}, where
the second coefficient is
$657/540$ instead of $-29/40$; in particular, it is not
positive definite contrary to the claim of \cite{hawking79}.

\section{Heat Traces on $S^3$}
\setcounter{equation}{0}

\subsection{Reduction to Irreducible Representations}

In this section we compute the effective action and the relevant
heat traces on the 3-sphere $S^3$ of radius $a$.
We define the dimensionless cosmological constant by
\be
\lambda=a^2\Lambda.
\ee
The curvature in the orthonormal frame has the form
\bea
R^{ab}{}_{cd} &=& \frac{1}{a^2}\eps^{fab}\eps_{fcd} = 
\frac{1}{a^2}\left(\d^a_c \d^b_d - \d^a_d \d^b_c\right), 
\\[5pt]
R_{ab} &=& \frac{2}{a^2}\d_{ab}, 
\\[5pt]
R &=& \frac{6}{a^2}\,.
\eea
Here and below $\varepsilon_{abc}$ is the three-dimensional Levi-Civita symbol.

The volume of the sphere is
\be
\vol(S^3)=2\pi^2a^3
\ee
and the
Euclidean classical Einstein-Hilbert action on $S^3$ is equal to
\be
S=\frac{\pi}{4G}a(-3+a^2\Lambda)\,.
\ee
Note that the classical action is bounded from below
and attains a minimum equal to
\be
S_0=-\frac{\pi}{2G\sqrt{\Lambda}}
\ee
at the radius determined by the cosmological constant
\be
a_0=\Lambda^{-1/2},
\ee
so that classically the dimensionless cosmological constant
is equal to $1$,
$
\lambda_0=1\,.
$
We will compute the heat trace of the Laplacian for the unit sphere $S^3$
by setting $a=1$; the trivial dimensional factor $a$ can be easily restored
at the end of the calculation by replacing $t\mapsto t/a^2$.

Let 
\be
(\Pi_0)^{ab}{}_{cd} = \frac{1}{3}g^{ab}g_{cd} 
\ee
be the projection to the scalar representation and
\be
\Pi_2=I_2-\Pi_0
\ee
be the projection onto the space of traceless
symmetric tensors.
Note that the projection $\Pi_2$ acts as identity in the subspace
of traceless symmetric tensors.
In three dimensions, the dimensions of these subspaces are
\be
\tr I_1=3, \qquad
\tr I_2= 6,
\ee
\be
\tr\Pi_0=1, \qquad
\tr \Pi_2=5.
\ee
This is consistent with the dimension of the general irreducible representation 
labeled by an integer $j$,
\be
\tr_j\Pi_j=2j+1\,.
\ee
Then the potential terms are
\bea
Q_1 &=& -2I_1\,,
\\
Q_2 &=&
 (4-2\lambda) \Pi_2-(2+2\lambda)\Pi_0\,.
\label{4.10xx}
\eea

We will reduce the calculation of the heat traces
of the operators $L_1$ and $L_2$ to the calculation of the
heat trace on the unit sphere $S^3$ of pure Laplacians $\Delta_j$ acting on irreducible
representations $j$,
\be
\Theta_j(t)=\Tr\exp(t\Delta_j).
\ee
First of all, we immediately see that 
since the endomorphism $Q_1$ is constant, we have
\be
\exp(-tQ_1) = e^{2t}I_1,
\ee
and, therefore,
the heat trace of the ghost operator is
\be
\Theta_{L_1}(t)=e^{2t}\Theta_1\left(t\right)\,.
\label{4.19xx}
\ee
We also have a similar formula for the operator $L_2$,
\be
\Theta_{L_2}(t)=\Tr\exp(-tQ_2)\exp(t\Delta)\,.
\ee

However, the general tensor representation contains
the irreducible representation with $j=2$ (traceless
symmetric two-tensors) and the scalar representation 
with $j=0$ (trace). The space of symmetric tensors
decomposes canonically into the direct sum
of the traceless tensors and scalars with the corresponding
projections $\Pi_2$ and $\Pi_0$.
It is easy to see that
\bea
\exp(-tQ_2) &=& e^{(-4+2\lambda)t}\Pi_2
+e^{(2+2\lambda)t}\Pi_0 .
\eea
Therefore, the heat trace of the graviton operator takes
the form
\bea
\Theta_{L_2}(t)
&=&e^{(-4+2\lambda)t}\Theta_2\left(t\right)
+e^{(2+2\lambda)t}\Theta_0\left(t\right).
\label{4.22xx}
\eea

\subsection{Heat Trace for Irreducible Representations}

Because the graviton operator neatly splits, we only need to compute the heat traces for Laplacians
in irreducible representation $j$ for {\it integer} $j$.
This heat trace can be computed
by using the heat kernel diagonal for the Laplacian $\Delta_j$
on the unit sphere $S^3$ given by the 
eqs. (6.16) of our paper \cite{avramidi12b}.
To get the heat trace we have to multiply the heat kernel diagonal by
the volume of the sphere $S^3$ equal to $\vol(S^3)=2\pi^2$
and by the dimension of the representation $j$ equal to $(2j+1)$.
This gives
\be
\Theta_{j}(t) = \frac{\sqrt{\pi}}{4}t^{-3/2}
e^{t[j(j+1)+1]}
\sum_{n = -\infty}^\infty 
\sum_{|\mu|\le j}
\exp\left( -\frac{\pi^2 n^2}{t}-\mu^2 t\right)
\left(1 - 2\mu^2 t-\frac{2\pi^2 n^2}{t}\right).
\ee

Following
\cite{avramidi12b} we introduce 
the function
\be
\Omega(t) = \sum_{n = -\infty}^\infty \exp\left(-\frac{n^2 \pi^2}{t}\right)\,,
\label{4.23xx}
\ee
which can be expressed in terms of the
Jacobi theta function
\be
\Omega(t)=\theta_3\left(0, e^{-\pi^2/t}\right),
\ee
and satisfies the following duality relation
\be
\Omega(t)=\sqrt{\frac{t}{\pi}}\Omega\left(\frac{\pi^2}{t}\right)
=\sqrt{\frac{t}{\pi}} \sum_{n = -\infty}^\infty e^{-tn^2}
=\sqrt{\frac{t}{\pi}}\theta_3\left(0, e^{-t}\right).
\ee

This enables us to express the function 
$\Theta_j$ in terms of the function $\Omega$ as follows
\bea
\Theta_{j}(t) &=& 
\frac{\sqrt{\pi}}{4}t^{-3/2}
e^{t[j(j+1)+1]}\sum_{|\mu|\le j} e^{-\mu^2 t}
\left[(1-2\mu^2 t)\Omega(t) 
- 2t\Omega'(t)\right].
\eea
By using the duality relation, the function $\Theta_j$ takes the form
\bea
\Theta_{j}(t) &=& 
e^{t[j(j+1)+1]}\sum_{n=-\infty}^\infty\sum_{|\mu|\le j} 
\frac{1}{2}\left(n^2-\mu^2\right)e^{-t(n^2+\mu^2)} \,.
\eea
Finally by using the obvious equation
\be
\sum_{\mu=-j}^j \sum_{n=-j}^j
\left(n^2-\mu^2\right)e^{-t(n^2+\mu^2)}=0 \,,
\ee
we get
the heat trace of pure Laplacian 
in the irreducible representation $j$
\bea
\Theta_j(t) &=&
\sum_{n=j+1}^\infty
\Biggl\{
n^2\exp\left\{-t\left[n^2-j(j+1)-1\right]\right\}
\nonumber\\
&&
+
\sum_{\mu=1}^j
2(n^2-\mu^2)\exp\left\{-t\left[n^2+\mu^2-j(j+1)-1\right]\right\}
\Biggr\}.
\label{4.57xx}
\eea

In particular, the eq. (\ref{4.57xx}) gives the eigenvalues
and their multiplicities 
of the pure Laplacian acting on an irreducible representation
$j$ of $SU(2)$. It is labeled by two integers $n$ and $\mu$ such that
\be
0\le\mu\le j<n.
\ee
The eigenvalues  are given by
\bea
\lambda_{n,\mu}(-\Delta_j) &=& n^2+\mu^2-j(j+1)-1,
\eea
and their multiplicities are
\be
d_{n,0}(-\Delta_j) = n^2,
\ee
for $\mu=0$ and
\be
d_{n,\mu}(-\Delta_j) = 2(n^2-\mu^2),
\ee
for $1\le \mu\le j$.

The minimal eigenvalue of the Laplacian $-\Delta_j$ is
\be
\lambda_{\rm min}(-\Delta_j)=j
\ee
with multiplicity $d_{\rm min}(-\Delta_j)=(j+1)^2$.
In particular, this means that all Laplacians $-\Delta_j$ for 
$j\ge 1$ are positive and the scalar Laplacian $\Delta_0$
is non-negative, it has the obvious constant zero mode.

We will need the functions $\Theta_0$, $\Theta_1$ and $\Theta_2$,
\bea
\Theta_{0}(t) &=& 
\frac{\sqrt{\pi}}{4}t^{-3/2}
e^{t}\left[\Omega(t) 
- 2t\Omega'(t)\right]
\nonumber\\
&=& \sum_{n=1}^\infty n^2 e^{-t(n^2-1)},
\\
\Theta_{1}(t) &=& 
\frac{\sqrt{\pi}}{4}t^{-3/2}
\left\{\left[e^{3t}+2(1-2t)e^{2t}\right]\Omega(t)
-2t\left[e^{3t}+2e^{2t}\right]\Omega'(t)
\right\}
\nonumber\\
&=&
\sum_{n=2}^\infty
\left\{
n^2e^{-t(n^2-3)}
+2\left(n^2-1\right)e^{-t(n^2-2)}
\right\},
\\
\Theta_{2}(t) &=& 
\frac{\sqrt{\pi}}{4}t^{-3/2}
\Biggl\{
\left[e^{7t}+2(1-2t)e^{6t}+2(1-8t)e^{3t}\right]\Omega(t)
\nonumber\\
&&
-2t\left[e^{7t}+2e^{6t}+2e^{3t}\right]\Omega'(t)
\Biggr\}
\nonumber\\
&=&
\sum_{n=3}^\infty
\left\{
n^2e^{-t(n^2-7)}
+2\left(n^2-1\right)e^{-t(n^2-6)}
+2(n^2-4)e^{-t(n^2-3)}
\right\}.
\eea
It is worth noting that the contribution of $\mu=0$ and $\mu=1$
for $j=1$ corresponds to the decomposition of the vector fields
\be
\varphi_\mu=A^\perp_\mu+\nabla_\mu \sigma,
\ee 
where $A_\mu$ is the transversal (divergence free) vector,
and the contribution of $\mu=0$, $\mu=1$ and $\mu=2$
for $j=2$ corresponds to the decomposition of the trace-free
symmetric tensor fields
\be
\varphi_{\mu\nu}=\varphi^\perp_{\mu\nu}
+2\nabla_{(\mu}A^\perp_{\nu)}+\nabla_\mu\nabla_\nu\sigma
-\frac{1}{3}g_{\mu\nu}\Delta\sigma,
\ee 
where $\varphi^\perp_{\mu\nu}$ is the transversal (divergence free)
tracefree tensor and $\sigma$ is a scalar.

\subsection{Heat Trace of Quantum Gravity}

We introduce the trace-free tensor part and the scalar part of the graviton operator $L_2$
\bea
L_2^{(0)} &=& \Pi_0L_2\Pi_0,
\\
L_2^{(2)} &=& \Pi_2 L_2\Pi_2\,.
\eea
Now, by using the eqs. (\ref{4.19xx}) and (\ref{4.22xx}) we
compute 
the eigenvalues of the operators $L_1$ and $L_2$
\begin{align}
\lambda_{n,\mu}({L_1})&= \lambda_{n,\mu}(-\Delta_1)-2=n^2+\mu^2-5,
\qquad n\ge 2, \qquad \mu=0,1,
\\[5pt]
\lambda_{n,\mu}({L^{(0)}_2})&= \lambda_{n,0}(-\Delta_0)-2-2\lambda=n^2-3-2\lambda,
\qquad n\ge 1, 
\\[5pt]
\lambda_{n,\mu}({L^{(2)}_2})&= \lambda_{n,\mu}(-\Delta_2)+4-2\lambda=n^2+\mu^2-3-2\lambda,
\qquad n\ge 3, \qquad \mu=0,1,2.
\end{align}
The minimal eigenvalues are
\bea
\lambda_{\rm min}(L_1) &=& -1,
\\
\lambda_{\rm min}(L_2^{(0)}) &=& -2-2\lambda,
\\
\lambda_{\rm min}(L_2^{(2)}) &=& 6-2\lambda.
\eea
Notice that the minimal eigenvalue of the ghost operator $L_1$ is always
negative, the minimal eigenvalue of the conformal sector of the
graviton operator $L^{(0)}_2$ is negative for
$\lambda>-1$
and the minimal eigenvalue of the graviton operator in the traceless
tensor part $L_2^{(2)}$ is negative for $\lambda>3$.
That is, the graviton operator is positive only for negative
cosmological constant when $\lambda<-1$.

Next, by using eq. 
(\ref{25xx}) and (\ref{4.22xx}) we get
\bea
\Theta_{GR}(t) 
&=&
e^{(-4+2\lambda)t}\Theta_{2}\left(t\right) 
+e^{(2+2\lambda)t}\Theta_{0}\left(t\right) 
-2e^{2t}\Theta_{1}\left(t\right).
\eea
We can write this either in terms of the function $\Omega$
\bea
\Theta_{GR}(t) 
&=&
\frac{\sqrt{\pi}}{4}t^{-3/2}
\Biggl\{
-2t\left[e^{2\lambda t}\left(2e^{3t}+2e^{2t}+2e^{-t}\right)
-2e^{5t}-4e^{4t} \right]\Omega'(t)
\\
&&
+\left[e^{2\lambda t}\left(2e^{3t}+2(1-2t)e^{2t}+2(1-8t)e^{-t}\right)
-2e^{5t}-4(1-2t)e^{4t} \right]\Omega(t) 
\Biggr\},
\nonumber
\eea
which is useful in the ultraviolet limit as $t\to 0$,
or in the spectral form
\bea
\Theta_{GR}(t) &=& 
e^{(2+2\lambda) t}
+4e^{(-1+2\lambda) t}
-8e^{t}
-12
\nonumber\\[5pt]
&&
+\sum_{n=3}^\infty
\Biggr\{
e^{2\lambda t}\left\{
2n^2e^{-t(n^2-3)}
+2\left(n^2-1\right)e^{-t(n^2-2)}
+2(n^2-4)e^{-t(n^2+1)}
\right\}
\nonumber\\[5pt]
&&
-2n^2e^{-t(n^2-5)}
-4\left(n^2-1\right)e^{-t(n^2-4)}
\Biggr\},
\eea
which is useful in the infrared limit as $t\to \infty$.

When the classical 
Einstein equations are satisfied, that is, when $\lambda=1$,
the heat trace simplifies to
\bea
\Theta_{GR}(t) &=& 
\frac{\sqrt{\pi}}{4}t^{-3/2}
\Biggl\{
\left[-2(1-2t)e^{4t}+2(1-8t)e^{t}
\right]\Omega(t) 
-2t\left[-2e^{4t}+2e^{t}
\right]\Omega'(t)
\Biggr\}
\nonumber\\
&=&
e^{4t}
-4e^{t}
-12+
\sum_{n=3}^\infty
\left\{
-2\left(n^2-1\right)e^{-t(n^2-4)}
+2(n^2-4)e^{-t(n^2-1)}
\right\}.
\nonumber\\
\eea

The infrared properties are described by the 
limit $t \to \infty$. By using the
spectral representation of the heat trace we immediately get
\be
\Theta_{GR}(t)= 
e^{(2+2\lambda) t}
+4e^{(-1+2\lambda) t}
-8e^{t}
-12
+O(e^{(2\lambda-6)t}).
\ee
The exponential growth of the heat trace indicates the
presence of the negative modes.

It is instructive to study the asymptotics
of the heat trace as $t\to 0$.
By using the asymptotics of the function $\Omega$
as $t\to 0$
\be
\Omega(t)\sim 1, \qquad \Omega'(t)\sim 0,
\ee
we obtain
\be
\Theta_{GR}(t)=\frac{\sqrt{\pi}}{4} t^{-3/2}
\left\{C_0+tC_1+t^2C_2+ O(t^3)\right\}\,,
\ee
where
\bea
C_0 &=& 0,
\\
C_1 &=& 2\pi^2(12\lambda - 30),
\\
C_2 &=& 2\pi^2(12\lambda^2 - 24\lambda -3).
\eea
These coefficients coincide with the coefficients $C_k$
given by the general formulas (\ref{3.61xx})-(\ref{3.63xx})
in three dimensions, $n=3$.
Notice the absence of the constant term here.
This is the feature of three-dimensional quantum Einstein gravity
since it does not have any dynamics, that is, the 
number of degrees of freedom is equal to zero.

\section{Heat Traces on $S^1\times S^3$}
\setcounter{equation}{0}


\subsection{Reduction of Heat Traces}

In this section we study Einstein quantum gravity
in the physical four-dimensional Einstein Universe.
Since we would like to study the thermal effects at the same time,
we consider the four-dimensional Riemannian manifold 
$M=S^1\times S^3$ with a circle $S^1$ of radius $a_1$
and a sphere $S^3$ of radius $a$. So, all indices in this section are
four-dimensional, that is, they run over $1,2,3,4$.

Let $h^a{}_b$ be the projection tensor on $S^3$ and $q^a{}_b$
be the projection to $S^1$ so that
\be
\delta^a{}_b=q^a{}_b+h^a{}_b,
\ee
and
\be
h^a{}_b h^b{}_c=h^a{}_c, \qquad
q^a{}_b q^b{}_c=q^a{}_c,\qquad
h^a{}_b q^b{}_c=0, 
\ee
\be
h^a{}_a=3, \qquad q^a{}_a=1\,.
\ee
Also, we introduce the Levi-Civita tensor $\varepsilon_{abc}$
on $S^3$ such that
\be
\varepsilon_{abc}q^a{}_d=0,
\ee
and
\bea
\varepsilon_{abc}\varepsilon^{def}&=&
6h^d{}_{[a}h^e{}_{b}h^f{}_{c]},
\\
\varepsilon_{abc}\varepsilon^{dec}&=&
2h^d{}_{[a}h^e{}_{b]},
\\
\varepsilon_{abc}\varepsilon^{dbc}&=&
2h^d{}_{a}.
\eea

Then 
the curvature is
\bea
R^{ab}{}_{cd} &=&\frac{1}{a^2}(h^{a}{}_{c}h^{b}{}_{d}
-h^{a}{}_{d}h^{b}{}_{c}),
\\[5pt]
R_{ab} &=& \frac{2}{a^2}h_{ab},
\\[5pt]
R &=& \frac{6}{a^2}\,.
\eea
The volume of the manifold $M=S^1\times S^3$ is
\be
\vol(M)=4\pi^3a_1a^3,
\ee
and the classical action is equal to
\be
S=\frac{\pi^2}{2G}a_1a(-3+a^2\Lambda).
\ee

Thus, the potential terms (\ref{2.30xx}), (\ref{2.31xx}),
of the operators $L_1$ and $L_2$ are
\bea
a^2(Q_1)^a{}_b &=& -2h^a{}_b\,,
\\[5pt]
a^2(Q_2)^{ab}{}_{cd} &=& 
(6-2\lambda)\delta^a{}_{(c}\delta^b{}_{d)}
-2h^{(a}{}_{(c}h^{b)}{}_{d)}
-h^{ab}h_{cd}
\nonumber\\
&&
-h^{ab}q_{cd}
-q^{ab}h_{cd}
-4q^{(a}{}_{(c}h^{b)}{}_{d)}
-3q^{ab}q_{cd}.
\eea

We need to compute the heat traces of the Laplace type
operators $L_j=-\Delta+Q_j$ (\ref{2.34xx}) on $M=S^1\times S^3$.
We note that since the potential terms are constant we have
\be
\exp(-tL_j)=\exp(-tQ_j)\exp(t\Delta),
\ee
and also
\be
\exp(t\Delta^{S^1\times S^3})=\exp(t\Delta^{S^1})\exp(t\Delta^{S^3}).
\ee
Therefore, the heat traces $\Theta_{L_j}(t)$ of the operators $L_j$ can be
computed as follows
\be
\Theta_{L_j}(t)=\Theta^{S^1}\left(\frac{t}{a_1^2}\right)
\Theta_{L_j}^{S^3}\left(\frac{t}{a^2}\right),
\label{5.17xx}
\ee
where
\be
\Theta^{S^1}(t)=\sqrt{\frac{\pi}{t}}\;\Omega(t),
\label{5.18xx}
\ee
with $\Omega(t)$ defined by (\ref{4.23xx}),
is the heat trace on the unit circle $S^1$ and
\be
\Theta^{S^3}_{L_j}(t)=\Tr\exp(-tL_j^{S^3})
\ee
is the heat trace on the unit sphere $S^3$.

The heat trace $\Theta^{S^3}_{L_j}(t)$ on $S^3$ was computed in our paper \cite{avramidi12b}.
We consider a tensor representation of spin $j$ of the spin group $\Spin(4)$
with generators $\Sigma^{ab}_{(j)}$ satisfying the algebra
(\ref{3.5xx}). Recall that $\Spin(4)=SU(2)\times SU(2)$.
Therefore, the matrices
\be
G_{(j)i}=\frac{1}{2}\varepsilon_{iab} \Sigma_{(j)}^{ab},
\ee
satisfy the algebra (no summation over $j$!)
\be
[G_{(j)i},G_{(j)k}]=-\varepsilon^l{}_{ik}G_{(j)l}
\ee
and form a {\it reducible} representation of the group $SU(2)$;
with the Casimir operator (no summation over $j$!)
\be
G_{(j)}^2=G_{(j)i}G_{(j)i}.
\ee
We also define the matrix
\be
G_{(j)}(y)=G_{(j)i}y^i,
\ee
where $y=(y^i)$ is a unit vector.
Let $f$ be a real-valued function of $x=(x^i)\in \RR^3$. Let $x^i=ry^i$, 
where $r=|x|=\sqrt{x^ix_i}$ and $y=(y^i)$ is
the unit vector such that $|y|=1$. 
Of course, the unit vector $y$ lies on the unit sphere $S^2$ in $\RR^3$.
We introduce the average over the unit sphere $S^2$ of functions in $\RR^3$ by
\be
\left<f\right>(r)=\frac{1}{4\pi} \int\limits_{S^2} dy_{S^2}f(ry);
\label{group_average}
\ee
the integration goes over the unit sphere $S^2$
with the appropriate induced
metric on $S^2$.

Then the heat trace of the Laplace type operator $L_j=-\Delta+Q_j$
on the unit sphere $S^3$
has the  form \cite{avramidi12b}
\be
\Theta^{S^3}_{L_j}(t)=\frac{\sqrt{\pi}}{4}\; t^{-3/2}
\tr\exp\left[-t(G_{(j)}^2+Q_j-I_j)\right]
S_j\left(t\right),
\label{5.25xx}
\ee
where $S_j(t)$
\bea
S_j(t)&=&
\sum_{n=-\infty}^\infty
\exp\left(-\frac{\pi^2 n^2}{t}\right)
\int\limits_{-\infty}^\infty \frac{dr}{\sqrt{\pi }} \;e^{-r^2}
\left(2r^2-2\frac{\pi^2 n^2}{t}\right)
\left<\exp\left[2r\sqrt{t}\,G_{(j)}(y)\right]\right>\,.
\nonumber\\
\label{614xx}
\eea

\subsection{Generators}

For the vector and the symmetric 2-tensor representation
(\ref{3.7xx}) and (\ref{3.16xx}) the generators have the form
\bea
(G_{(1)i})^c{}_d &=& \varepsilon_i{}^c{}_d\,,
\\
(G_{(2)i})^{ef}{}_{cd} &=& 2\varepsilon_i{}^{(e}{}_{(c}\delta^{f)}{}_{d)}\,,
\eea
so that
\bea
(G_{(1)}(y))^a{}_b &=& \eps_i{}^a{}_b y^i \,, 
\\
(G_{(2)}(y))^{ab}{}_{cd} &=& 2\eps_i{}^{(a}{}_{(c}\delta^{b)}{}_{d)} y^i \,.
\eea
We compute the Casimir operators
\bea
(G_{(1)}^2)^c{}_d
&=&-2h^c{}_d,
\\
(G_{(2)}^2)^{ef}{}_{cd}
&=&
-6h^{(e}{}_{(c}h^{f)}{}_{d)}
+2h^{ef}h_{cd}
-4q^{(e}{}_{(c}h^{f)}{}_{d)}
\eea
and the sums
\bea
(G_{(1)}^2+Q_1)^c{}_d &=& -4h^c{}_d,
\\[5pt]
(G_{(2)}^2+Q_2)^{ab}{}_{cd}
&=&
(6-2\lambda)\delta^a{}_{(c}\delta^b{}_{d)}
-8h^{(a}{}_{(c}h^{(b)}{}_{d)}
+h^{ab}h_{cd}
\nonumber\\
&&
-h^{ab}q_{cd}
-q^{ab}h_{cd}
-8q^{(a}{}_{(c}h^{(b)}{}_{d)}
-3q^{ab}q_{cd}.
\eea

\subsection{Algebra of Constant Symmetric Endomorphisms}

First of all, for the vector representation 
we immediately obtain
\begin{lemma}
\be
\exp\left\{-t(G_{(1)}^2+Q_1)\right\}
=I_1-H+e^{4t}H\,,
\label{5.36xx}
\ee
where $H$ is the matrix of the projection $H=(h^a{}_b)$.
\end{lemma}

To compute this exponential for the tensor representation
we need to do some algebra.
We define the following basis of endomorphisms
acting on symmetric two-tensors in four dimensions
\bea
I^{ab}{}_{cd} &=& \delta^{(a}{}_{(c}\delta^{b)}{}_{d)},
\\
A^{ab}{}_{cd} &=& h^{(a}{}_{(c}h^{b)}{}_{d)},
\\
B^{ab}{}_{cd} &=& h^{ab}h_{cd},
\\
C^{ab}{}_{cd} &=& h^{ab}q_{cd},
\\
D^{ab}{}_{cd} &=& q^{ab}h_{cd},
\\
E^{ab}{}_{cd} &=& q^{(a}{}_{(c}h^{b)}{}_{d)},
\\
F^{ab}{}_{cd} &=& q^{ab}q_{cd}.
\eea

First, we note the identity
\be
A+2E+F=I\,,
\ee
so that
\be
E=\frac{1}{2}(I-A-F).
\ee
Of course, $I$ is the identity. We compute the squares
of these endomorphisms
\bea
A^2 &=& A,
\\
B^2&=& 3B,
\\
C^2&=&0,
\\
D^2&=&0,
\\
E^2&=&\frac{1}{2}E,
\\
F^2&=&F,
\eea
and their products
\bea
AB &=&B,\qquad 
BA=B,
\\
AC &=&C,\qquad
CA=0,
\\
AD &=&0,\qquad
DA=D,
\\
AE &=&0,\qquad
EA=0,
\\
AF &=&0,\qquad
FA=0,
\\
BC &=&3C,\qquad
CB=0,
\\
BD &=&0,\qquad
DB=3D,
\\
BE &=&0,\qquad
EB=0,
\\
BF &=&0,\qquad
FB=0,
\\
CD &=&B,\qquad
DC=3F,
\\
CE &=&0,\qquad
EC=0,
\\
CF &=&C,\qquad
FC=0,
\\
DE &=&0,\qquad
ED=0,
\\
DF &=&0,\qquad
FD=D
\\
EF &=&0,\qquad
FE=0.
\eea

Next, we define the following endomorphisms
\bea
P_1&=& A-\frac{1}{3}B,
\\[5pt]
P_2&=& \frac{1}{3}B,
\\[5pt]
P_3&=& 2E,
\\[5pt]
P_4&=& F,
\\
T&=&C+D,
\\
X&=&\frac{1}{2}\left(P_4-P_2-T\right),
\\
\Pi_{\pm}&=&\frac{1}{2}\left(P_2+P_4\pm X\right).
\eea

By using the algebra of these endomorphisms one can prove
\begin{lemma}
The endomorphisms $P_1, P_2, P_3$ and $P_4$
form a set of orthogonal
projections satisfying
\be
P_i^2=P_i,
\label{5.40xx}
\ee
\be
P_iP_j=0, \qquad \mbox{if}\qquad i\ne j,
\label{5.41xx}
\ee
and
\be
P_1+P_2+P_3+P_4=I\,.
\label{5.42xx}
\ee
The dimensions of the corresponding subspaces
are determined by the traces
\bea
\tr P_1=5, \qquad
\tr P_2=1, \qquad
\tr P_3=3, \qquad
\tr P_4=1\,.
\eea
\end{lemma}

Of course, the total dimension of the space of symmetric two-tensors
in four dimensions is
\be
5+1+3+1=10\,.
\ee

\begin{lemma}
\begin{enumerate}
\item
The endomorphism $X$ satisfies the equations
\be
XP_1=P_1X=P_3X=XP_3=0\,.
\ee
\be
(P_2+P_4) X=X(P_2+P_4)=X\,.
\ee
\be
X^2=P_2+P_4.
\label{5.70xx}
\ee
\be
\tr X=0,
\ee
It has the eigenvalue $0$ with multiplicity $8$ and simple eigenvalues $-1,+1$.
\item
The endomorphisms $\Pi_\pm$ are the projections to the eigenspaces of $X$
corresponding to the eigenvalues $\pm 1$. They satisfy the equations
\be
\Pi_\pm^2=\Pi_\pm,
\ee
\be
\Pi_-\Pi_+=\Pi_+\Pi_-=0,
\ee
\be
X\Pi_\pm=\pm\Pi_\pm,
\label{5.74xx}
\ee
\bea
\Pi_+(P_2+P_4) = \Pi_+,
\qquad
\Pi_-(P_2+P_4) =\Pi_-.
\eea
\be
\tr \Pi_\pm = 1,
\ee
\end{enumerate}
\end{lemma}

{\it Proof.}
The projections $P_i$ act on the matrices $C$ and $D$ by
\bea
P_1C&=&0, \qquad CP_1= 0,\\
P_2C&=&C, \qquad CP_2= 0,\\
P_3C&=&0, \qquad CP_3= 0,\\
P_4C&=&0, \qquad CP_4=C,
\\
P_1D&=&0, \qquad DP_1= 0,\\
P_2D&=&0, \qquad DP_2= D,\\
P_3D&=&0, \qquad DP_3= 0,\\
P_4D&=&D, \qquad DP_4=0.
\eea
and, therefore,
\bea
P_1T&=&0, \qquad TP_1= 0,\\
P_2T&=&C, \qquad TP_2= D,\\
P_3T&=&0, \qquad TP_3= 0,\\
P_4T&=&D, \qquad TP_4= C,
\eea
%
%
%
so that
\be
P_2T+TP_2=T, \qquad
P_4T+TP_4=T.
\ee
Also, we have 
\be
T^2=3(P_2+P_4).
\ee
By using these equations one can prove all the equations of the lemma.

Since the matrix $X$ is orthogonal to the projections $P_1$ and $P_3$, 
it has an obvious eigenvalue equal to zero
with multiplicity $8=5+3$
making it essentially two-dimensional. It acts nontrivially
only on subspaces spanned by projections $P_2$ and $P_4$,
which are both one-dimensional.
Since it is obviously traceless,
 the sum of its eigenvalues is equal to zero.
 It is easy to see that it has two non-zero eigenvalues $\pm 1$.
This follows from the eqs. (\ref{5.70xx}).

The matrices 
$
\Pi_{\pm}
$
are the eigenprojections corresponding to the eigenvalues
$\pm 1$; this follows from the eqs. (\ref{5.74xx}).

We prove the following
\begin{lemma}
\be
\exp\left\{-t[G_{(2)}^2+Q_2]\right\}
=e^{2\lambda t}
\left\{P_1 e^{2t}
+P_3 e^{-2t}
+\Pi_- 
+\Pi_+e^{-4t}
\right\}.
\label{5.92xx}
\ee
\end{lemma}

{\it Proof.}
We have
\bea
G_{(2)}^2+Q_2 
&=&
\left(2-2\lambda\right)I
-4P_1+2X,
\eea
Therefore,
\be
\exp\left\{-t[G_{(2)}^2+Q_2]\right\}
=\exp\left\{-t\left(2-2\lambda\right)\right\}
\exp\left(4tP_1\right)
\exp(-2tX)
\ee
We compute
\bea
\exp\left(4tP_1\right)
&=&
P_1 e^{4t}+P_2+P_3+P_4\,.
\eea
The only thing left to compute is the exponential $\exp(-2tX)$.
By using
\be
X^{2n}=P_2+P_4, \qquad
X^{2n+1}=X,
\ee
we get
\bea
\exp(-2tX) 
&=&
P_1+P_3+\Pi_- e^{2t}
+\Pi_+e^{-2t}.
\eea
This finally gives the eq. (\ref{5.92xx}).

\subsection{Algebra of Symmetric Endomorphisms on $S^3$}

Let  $y=(y^i)$ be a unit vector orthogonal to $q^a{}_b$, that is, satisfying
$y^a=h^a{}_by^b$.
We introduce two matrices
\be
Z^a{}_b=y^i\varepsilon_i{}^a{}_b\,,
\ee
and
\be
P^a{}_b=h^a{}_b-y^ay_b\,.
\ee
The square of the matrix $Z$ is equal to
\be
Z^2=-P,
\label{5.100xx}
\ee
and the matrix $P$ is obviously a projection 
so that
\be
P^2=P,
\ee
\bea
PZ &=&ZP=Z,
\\
PH &=&HP=P,
\eea
and
\be
\tr Z=0, \qquad
\tr P=2,
\qquad
\tr H=3.
\ee

First, we prove 
\begin{lemma}
The exponential of the matrix $G_{(1)}(y)$ is
%
%
%
\be
\exp[2rG_{(1)}(y)]=I_1-P+\cos(2r) P+\sin(2r)Z
\label{5.115xx}
\ee
with the trace
%
%
%
\be
\tr\exp(2rG_{(1)}(y))=2+2\cos(2r).
\ee
\end{lemma}

{\it Proof.} This follows from the fact that
\be
G_{(1)}(y)=Z
\ee
and the eq. (\ref{5.100xx}).

Next, we introduce the following endomorphisms
acting on symmetric two-tensors
\bea
K^{ab}{}_{cd} &=& Z^{(a}{}_{(c}\delta^{b)}{}_{d)},
\\
L^{ab}{}_{cd} &=& Z^{(a}{}_{(c}Z^{b)}{}_{d)},
\\
W^{ab}{}_{cd} &=& Z^{(a}{}_{(c}P^{b)}{}_{d)},
\\
M^{ab}{}_{cd} &=&P^{(a}{}_{(c}\delta^{b)}{}_{d)},
\\
N^{ab}{}_{cd} &=& P^{(a}{}_{(c}P^{b)}{}_{d)},
\\
S^{ab}{}_{cd} &=& P^{ab}P_{cd},
\\
U^{ab}{}_{cd} &=& P^{ab}g_{cd},
\\
Y^{ab}{}_{cd} &=& g^{ab}P_{cd}.
\eea
We compute the traces
\bea
\tr K &=& 0,
\\
\tr L &=& -1,
\\
\tr M &=& 5,
\\
\tr N &=& 3,
\\
\tr W &=& 0.
\eea
We need to compute the algebra of these endomorphisms.
First, we have
\bea
K^2 &=&-\frac{1}{2}M+\frac{1}{2}L,
\\
M^2 &=&M,
\\
L^2 &=&N,
\\
N^2 &=&N,
\\
LM &=&L,
\\
ML &=&L,
\\
KM &=&\frac{1}{2}(K+W),
\\
KL &=&-W,
\\
NM &=&N,
\\
NL &=&L,
\\
KN&=& W,
\\
KW&=& \frac{1}{2}(-N+L).
\eea

We prove the following
\begin{lemma}
The exponential of the endomorphism $G_{(2)}(y)$ has the form
\be
\exp[2r G_{(2)}(y)]=\gamma(r) I+\mu(r) M+\nu(r) N+\lambda(r) L+\eta(r) W+\varkappa(r) K,
\label{5.143xx}
\ee
where
\bea
\gamma(r)&=&1,
\label{5.130xx}
\\[5pt]
\mu(r) &=& 2\cos(2r)-2,
\\[5pt]
\nu(r) &=& \frac{1}{2}\cos(4r)-2\cos(2r)+\frac{3}{2},
\\[5pt]
\lambda(r) &=& \frac{1}{2}-\frac{1}{2}\cos (4r),
\\[5pt]
\eta(r) &=& \sin(4r)-2\sin(2r),
\\[5pt]
\varkappa(r) &=& 2\sin(2r).
\label{5.135xx}
\eea
with the trace
\bea
\tr \exp[2rG_{(2)}(y)] 
&=&
4+4\cos(2r)+2\cos(4r).
\eea

\end{lemma}

{\it Proof.}
We note that
\be
G_{(2)}(y)=2K.
\ee
Let
\be
J(r)=\exp(4rK).
\ee
It satisfies the differential equation
\be
\partial_r J=4KJ
\ee
with initial condition
\be
J(0)=I.
\ee
We decompose it according to
\be
J=\gamma I+\mu M+\nu N+\lambda L+\eta W+\varkappa K.
\ee
Then by using the algebra of the matrices  $M, N, L, W, K$
we have
\be
KJ=-\frac{\varkappa}{2}M-\frac{\eta}{2}N
+\frac{1}{2}\left(\eta+\varkappa\right)L
+\left(\frac{\mu}{2}+\nu-\lambda\right)W
+\left(\frac{\mu}{2}+\gamma\right)K.
\ee
Therefore, the coefficients of this expansion 
must satisfy the differential equations
\bea
\partial_r\gamma &=&0,
\\
\partial_r\mu &=&-2\varkappa,
\\
\partial_r\nu &=&-2\eta,
\\
\partial_r\lambda &=&2\eta+2\varkappa,
\\
\partial_r\eta &=&2\mu +4\nu-4\lambda,
\\
\partial_r\varkappa &=&2\mu+4\gamma,
\eea
with the initial conditions
\be
\gamma(0)=1, \qquad
\mu(0)=\nu(0)=\lambda(0)=\eta(0)=\varkappa(0)=0\,.
\ee
The solution of this system gives the result (\ref{5.130xx})-(\ref{5.135xx}).
Now the trace can be easily computed.

\subsection{Group Averages}

Next, we need to compute the group averages  (\ref{group_average}) of the functions
given by (\ref{5.115xx}) and (\ref{5.143xx}).
Thus, we need to compute the averages of the polynomials.
We prove
\begin{lemma}
The averages of the monomials are
\bea
\left<1\right>&=&1,
\\
\left<y^{i_1}\cdots y^{i_{2k+1}}\right>&=&0,
\\
\left<y^{i_1}\cdots y^{i_{2k}}\right>&=&
\frac{1}{2k+1}\delta^{(i_1i_2}\cdots \delta^{i_{2k-1}i_{2k})}.
\label{5.156xx}
\eea
\end{lemma}

{\it Proof.} The first two equations are obvious.
To prove the eq. (\ref{5.156xx}) we consider the
Gaussian integral
\bea
\int\limits_{\RR^3} dx \; e^{-|x|^2}
x^{i_1}\cdots x^{i_{2n}}
&=&
\pi^{3/2}\frac{(2n)!}{n!2^{2n}}\delta^{(i_1i_2}\cdots \delta^{i_{2n-1}i_{2n})}.
\eea
By changing the variables here by $x^i=ry^i$ and using the integral
\be
\int\limits_0^\infty dr\; r^{2n+2} e^{-r^2}
=\sqrt{\pi}\frac{(2n+2)!}{(n+1)!2^{2n+3}}
\ee
we get eq. (\ref{5.156xx}).

\begin{corollary}
Let $h$ be the projection onto the three-dimensional
subspace $V=\RR^3$ of $\RR^4$ and $y=(y^i)$ be a four-dimensional 
unit vector lying in $V$.
Then the eq. (\ref{5.156xx}) is modified as follows
\bea
\left<y^{a_1}\cdots y^{a_{2k}}\right>&=&
\frac{1}{2k+1}h^{(a_1i_2}\cdots h^{a_{2k-1}a_{2k})}.
\label{5.159xx}
\eea
\end{corollary}


We define the characters of an irreducible representation $j$ of $SU(2)$ by
\be
\chi_j(r) = \tr_j\left<\exp[2rG_{(j)}(y)]\right>\,,
\ee
where $\tr_j$ is the trace in the irreducible representation $j$.
For an irreducible representation $j$ the average of a group
element over the $S^2$ is given by eq. (5.55) of \cite{avramidi12b}:
\be
\left<\exp[2rG_{(j)}(y)]\right> = \frac{1}{2j+1}\sum_{|\mu|\le j} 
\cos(2\m r) \Pi_j,
\ee
so that
\be
\chi_j(r) = \sum_{|\mu|\le j} \cos(2\m r).
\label{332xx}
\ee
in particular,
\bea
\chi_{1}(r) &=& 1 + 2\cos(2r)\,, \\
\chi_{2}(r) &=& 1+ 2\cos(2r) + 2\cos(4r)\,.
\label{334xx}
\eea


Using the
averages of the monomials calculated above we obtain
\bea
\left<Z\right>&=&0,
\\
\left<P\right>&=&=\frac{2}{3}h,
\\
\left<M\right> &=& 
 \frac{2}{3}\left(P_1 + P_2 + \frac{1}{2}P_3 \right),
\\[5pt]
\left<N\right> &=&  
\frac{7}{15}P_1 + \frac{2}{3}P_2,
\\[5pt]
\left<L\right> &=& 
\frac{1}{3}(2P_2 - P_1),
\\[5pt]
\left<W\right> &=& 0,
\\[5pt]
\left<K\right> &=& 0.
\eea
This allows us to calculate the group averages of the exponentials
%
%
%
\bea
\left<\exp[2rG_{(1)}(y)]\right> &=& 
I_1-\frac{2}{3}H+\frac{2}{3}H\cos(2r),
\\
\left<\exp[2rG_{(2)}(y)]\right> &=& \frac{1}{5}\left[1 + 2\cos(2r)+2\cos(4r)\right]P_1 + P_2 + P_4 
\nonumber\\
&&
+ \frac{1}{3}\left[1+2\cos(2r)\right]P_3.
\eea

\subsection{Heat Trace of Operator $L_1$ on $S^3$}


To compute the functions $S_j$ we will need the integrals
\bea
\int\limits_{-\infty}^\infty dr\,e^{-r^2}\cos\left(2\mu\sqrt{t}r\right) &=&
\sqrt{\pi}\,e^{-t\mu^2},
\label{5.184xx}
\\
\int\limits_{-\infty}^\infty dr\, e^{-r^2} \cos\left(2\mu\sqrt{t}r\right) r^2 &=&
\sqrt{\pi}\left(\frac{1}{2}-\mu^2 t\right)e^{-t\mu^2}.
\label{5.185xx}
\eea
By using these integrals we obtain
\be
\int^\infty_{-\infty}dr\; e^{-r^2}\left(2r^2 - \frac{2\pi^2n^2}{t} \right)\cos(2r\sqrt{t})
= \sqrt{\pi}\left(1-2t-\frac{2\pi^2n^2}{t}\right)e^{-t},
\ee
and we finally obtain from (\ref{614xx})
%
%
%
\be
S_{L_1}(t) =
\sum_{n=-\infty}^\infty
\exp\left(-\frac{\pi^2 n^2}{t}\right)
\Bigg[
\left(1 - \frac{2\pi^2n^2}{t}\right)\left(I_1-\frac{2}{3}H\right) 
+ \left(1 - 2t -\frac{2\pi^2n^2}{t}\right)e^{-t} 
\frac{2}{3}H
\Bigg].
\ee

The heat trace is calculated now by using (\ref{5.25xx}) and (\ref{5.36xx})
%
%
%
\bea
\Theta^{S^3}_{L_1}(t) &=& \frac{\sqrt{\pi}}{4}t^{-3/2}
e^{t}\sum_{n=-\infty}^\infty \exp\left(-\frac{\pi^2 n^2}{t}\right)
\nonumber\\
&&\times
\tr\Bigg\{\left(1-\frac{2\pi^2n^2}{t} \right)\left(I_1-H+\frac{1}{3}He^{4t}
\right) 
\nonumber\\
&&
+\left(1-2t-\frac{2\pi^2n^2}{t} \right)e^{3t}
\frac{2}{3}H\Bigg\}.
\nonumber\\
&=& \frac{\sqrt{\pi}}{4}t^{-3/2}
\sum_{n=-\infty}^\infty \exp\left(-\frac{\pi^2 n^2}{t}\right) 
\\
&&
\times \Bigg\{\left(1-\frac{2\pi^2n^2}{t} \right)(e^{t}+e^{5t}) 
+2\left(1-2t-\frac{2\pi^2n^2}{t} 
\right)e^{4t}\Bigg\}.
\nonumber
\eea

The heat trace can be written in terms of the function $\Omega$,
\bea
\Theta^{S^3}_{L_1}(t)&=&
\frac{\sqrt{\pi}}{4}t^{-3/2}
\Bigg\{
\left[e^{5t}+e^t+2(1-2t)e^{4t}\right]\Omega(t)
-2t\left[e^{5t}+e^t+2e^{4t}\right]\Omega'(t)
\Bigg\}.
\nonumber\\
\eea
This heat trace has the asymptotic expansion as $t \to 0$
\be
\Theta^{S^3}_{L_1}(t)
=\frac{\sqrt{\pi}}{4}t^{-3/2}
\left(4+10t+13t^2+O(t^3)\right).
\label{5.190xx}
\ee

This can be put in the spectral form by using the identities 
\bea
\Omega(t)-2t\Omega'(t) &=& \frac{2t^{3/2}}{\sqrt{\pi}}
\sum_{n \in \ZZ}n^2e^{-tn^2},
\label{5.165xx}
\\
(1-2t)\Omega(t)-2t\Omega'(t) 
&=& \frac{2t^{3/2}}{\sqrt{\pi}}\sum_{n \in \ZZ}(n^2-1)e^{-tn^2}.
\label{5.167xx}
\eea
We get
\be
\Theta^{S^3}_{L_1}(t)=\frac{1}{2}
\sum_{n \in \ZZ}e^{-tn^2}
\left[n^2\left(e^{5t}+e^{t}\right)+2(n^2-1)e^{4t}\right].
\ee
The asymptotic behavior of the heat trace in the limit $t \to \infty$ is
\be
\Theta^{S^3}_{L_1}(t) =4e^{t} + 7+O(e^{-4t})\,.
\ee

\subsection{Heat Trace of the Operator $L_2$ on $S^3$}

Now, by using the integrals (\ref{5.184xx}) and (\ref{5.185xx}), we compute first
\bea
\int^\infty_{-\infty}dr\; e^{-r^2}\left(2r^2 - \frac{2\pi^2n^2}{t} \right)
&=& \sqrt{\pi}\left(1-\frac{2\pi^2n^2}{t}\right), 
\\
\int^\infty_{-\infty}dr\; e^{-r^2}\left(2r^2 - \frac{2\pi^2n^2}{t} \right)\cos(4r\sqrt{t})
&=& \sqrt{\pi}\left(1-8t-\frac{2\pi^2n^2}{t}\right)e^{-4t}.
\eea
and then by using (\ref{614xx}) we find
\bea
S_2(t)&=&
\sum_{n=-\infty}^\infty
\exp\left(-\frac{\pi^2 n^2}{t}\right)
\Bigg[
\left(\frac{1}{5}P_1 + P_2 + \frac{1}{3}P_3 + P_4\right)\left(1-\frac{2\pi^2n^2}{t}  \right) \\
&&
+ \left(\frac{2}{5}P_1 + \frac{2}{3}P_3\right)\left(1-2t-\frac{2\pi^2n^2}{t}  \right)e^{-t}
+ \frac{2}{5}P_1\left(1-8t-\frac{2\pi^2n^2}{t}  \right)e^{-4t} \nonumber
\Bigg].
\eea
Further, by using (\ref{5.92xx}) and the algebra of symmetric endomorphisms we get
\bea
&& \exp[-t(G_{(2)}^2 + Q_2^2)]S_2(t) =  
e^{2\lambda t}\sum_{n=-\infty}^\infty
\exp\left(-\frac{\pi^2 n^2}{t}\right) 
\\
&&\times
\Bigg\{
\left[\frac{1}{5}\left(1 -\frac{2\pi^2n^2}{t}  \right)e^{2t}
+ \frac{2}{5}\left(1-8t -\frac{2\pi^2n^2}{t}  \right)e^{-2t} 
+ \frac{2}{5}\left(1-2t -\frac{2\pi^2n^2}{t}  \right)e^{t}\right]P_1 
\nonumber \\
&&
+\left[\frac{1}{3}\left(1 -\frac{2\pi^2n^2}{t}  \right)e^{-2t} 
+ \frac{2}{3}\left(1-2t -\frac{2\pi^2n^2}{t}  \right)e^{-3t}\right]P_3
+ \left(\Pi_- + \Pi_+ e^{-4t}\right)\left(1-\frac{2\pi^2n^2}{t}  \right) 
\nonumber
\Bigg\}.
\eea
Finally, by taking the trace we obtain from eq. (\ref{5.25xx})
the heat trace
\bea
\Theta^{S^3}_{L_2}(t) &=& \frac{\sqrt{\pi}}{4}t^{-3/2}
e^{2\lambda t}\sum_{n=-\infty}^\infty \exp\left(-\frac{\pi^2 n^2}{t}\right) 
\nonumber \\
&& \times
\Bigg\{\left(1-\frac{2\pi^2n^2}{t} \right)e^{3t} 
+ 2\left(1-2t-\frac{2\pi^2n^2}{t} \right)e^{2t} 
\nonumber \\
&&
+\left(1-\frac{2\pi^2n^2}{t} \right)e^{t} 
+\left(3-16t-\frac{6\pi^2n^2}{t} \right)e^{-t} 
\nonumber \\
&&
+ 2\left(1-2t-\frac{2\pi^2n^2}{t} \right)e^{-2t} 
+\left(1-\frac{2\pi^2n^2}{t} \right)e^{-3t})  \Bigg\}.
\eea
We can rewrite this in terms of the function $\Omega$ as
\bea
\Theta^{S^3}_{L_2}(t) 
&=& \frac{\sqrt{\pi}}{4}t^{-3/2}e^{2\lambda t}\Bigg\{
\Bigl[e^{3t}+2(1-2t)e^{2t}+e^{t}+(3-16t)e^{-t}+2(1-2t)e^{-2t}
\nonumber\\
&&
+e^{-3t}
\Bigr]\Omega(t)
-2t\left[e^{3t}+2e^{2t}+e^t+3e^{-t}+2e^{-2t}+e^{-3t}
\right]\Omega'(t)
\Bigg\}.
\eea
To second order in $t$, the exponentials may be expanded
\be
\Theta^{S^3}_{L_2}(t) =\frac{\sqrt{\pi}}{4}t^{-3/2}
\left[10 +(-26 + 20\lambda)t
 + (35-52\lambda+20\lambda^2)t^2+O(t^3)\right].
\label{5.201xx}
\ee 

By using the identities
(\ref{5.165xx})-(\ref{5.167xx}) and
\bea
(3-16t)\Omega(t)-6t\Omega'(t) 
&=& \frac{2t^{3/2}}{\sqrt{\pi}}\sum_{n \in \ZZ}(3n^2-8)e^{-tn^2}\,,
\eea
the heat trace can be rewritten in the spectral form
\bea
\Theta^{S^3}_{L_2}(t) &=& \frac{1}{2}e^{2\lambda t}
\sum_{n \in \ZZ}e^{-tn^2}
 \Bigg\{n^2e^{3t} +2(n^2-1)e^{2t}+n^2e^{t}
+(3n^2-8)e^{-t}
\nonumber\\
&&
+2(n^2-1)e^{-2t}+n^2e^{-3t} 
\nonumber
\Bigg\}.
\eea
The asymptotic behavior of the heat trace in the limit $t \to \infty$ is
\be
\Theta^{S^3}_{L_2}(t) = e^{2\lambda t}\left[1+O(e^{-3t})\right].
\ee

\section{Effective Action}
\setcounter{equation}{0}

Now, by using (\ref{5.17xx}), (\ref{5.18xx}), the heat trace of quantum gravity (\ref{25xx}) on $S^1\times S^3$
takes the form
\bea
\Theta_{GR}(t) &=&
a_1\sqrt{\frac{\pi}{t}}\Omega\left(\frac{t}{a^2_1}\right)
\left\{\Theta^{S^3}_{L_2}\left(\frac{t}{a^2}\right)
-2\Theta^{S^3}_{L_1}\left(\frac{t}{a^2}\right)
\right\}.
\eea
It will be convenient to separate the asymptotic behavior at $t\to 0$
\bea
\Theta_{GR}(t)
&=&
\frac{\pi}{4}\frac{a_1a^3}{t^2}
\Omega\left(\frac{t}{a^2_1}\right)W\left(\frac{t}{a^2}\right),
\eea
where
\be
W(t)=\frac{4}{\sqrt{\pi}}t^{3/2}
\left\{\Theta^{S^3}_{L_2}\left(t\right)
-2\Theta^{S^3}_{L_1}\left(t\right)
\right\}.
\ee
Note that the function $W$ depends also on the radius $a$ through the dimensionless
cosmological constant $\lambda=a^2\Lambda$.
The asymptotics of the function $W$ as $t\to 0$ are 
\bea
W(t) &=&c_0+c_1t+c_2t^2+O(t^3),
\eea
where the coefficients $c_k$ are computed from eqs. (\ref{5.190xx}) and (\ref{5.201xx})
\bea
c_0&=&2,
\\
c_1&=&-46+20\lambda,
\\
c_2&=&9-52\lambda+20\lambda^2.
\eea
The coefficients $c_k$ differ from the coefficients $C_k$, (\ref{2.35xx}), by the volume factor
$\vol(S^1\times S^3)=4\pi^3a_1a^3$ and a uniform factor $a^{2k}$. 
The asymptotics of the function $W$ as $t\to \infty$ are
\bea
W(t)=\frac{4}{\sqrt{\pi}}t^{3/2}\left\{
e^{2\lambda t}-8e^t-14+O(e^{-4t})+O(e^{(2\lambda-3)t})
\right\}\,.
\eea

Now, following \cite{avramidi12b}
the one-loop effective action can be presented in the form
\be
\Gamma_{(1)}=
-\frac{\pi}{8}\frac{a_1}{a}\left\{\beta\log\frac{\mu^2}{\mu_0^2}
+\Phi\right\},
\ee
where
\bea
\Phi &=&a^4\int\limits_0^\infty \frac{dt}{t^3}e^{-tz^2}
\left\{\Omega\left(\frac{t}{a^2_1}\right)W\left(\frac{t}{a^2}\right)
-R_{GR}\left(\frac{t}{a^2}\right)\right\},
\\
R_{GR}\left(t\right) &=& e^{-t\mu_0^2}\left\{
2+\left(c_1+2\mu_0^2\right) t
+\left(c_2+c_1\mu_0^2+\mu_0^4\right) t^2\right\},
\\[5pt]
\beta &=& c_2 - z^2a^2 c_1 +z^4a^4,
\eea 
$z$ is an infrared regularization parameter,
and $\mu_0$ is an arbitrary renormalization parameter.

The total effective action including the classical term 
in the one-loop approximation is 
\be
\Gamma =
\frac{\pi^2}{2G}a_1a(-3+\lambda)
-\hbar\frac{\pi}{8}\frac{a_1}{a}\left\{
\beta\log\frac{\mu^2}{\mu_0^2}
+\Phi\right\}+O(\hbar^2).
\ee
We neglect the terms of order $\hbar^2$ and set $\hbar=1$.

\section{Thermodynamics}
\setcounter{equation}{0}

The effective action is a function of two variables, $\Gamma=\Gamma(a_1,a)$,
where $a_1$ is the radius of the circle $S^1$ and $a$ is the radius of the $3$-sphere $S^3$.
The temperature $T$ is determined by the radius of the circle $a_1$ by
$T=1/(2\pi a_1)$ 
and the spatial volume $V$ of the system is the volume of the
sphere $S^3$, equal to $V=2\pi^2 a^3$.
We introduce a dimensionless temperature
\be
x=\frac{a}{a_1},
\ee
so that the the temperature is
\be
T=\frac{x}{2\pi a}.
\ee
Then for a canonical statistical ensemble with fixed $T$ and $V$ the free energy
$F$
is determined by the effective action $\Gamma$ by 
\be
F=T\Gamma=\frac{x}{2\pi a}\Gamma.
\ee
By using the results of the previous section 
for the effective action we obtain
the free energy
\be
F = 
\frac{\pi}{4G}a(-3+\lambda)
-\frac{1}{16a}\left(\beta\log\frac{\mu^2}{\mu_0^2}
+\Phi(x,a)\right)\,.
\label{74cc}
\ee

This enables one to compute all other thermodynamic parameters
of the graviton gas
such as the entropy 
\be
S= -\frac{\partial F}{\partial T}=-2\pi a \partial_x F,
\ee
the energy 
\be
E=F+TS=F-x\partial_x F,
\ee
the pressure 
\be
P=-\frac{\partial F}{\partial V}
=-\frac{1}{6\pi^2 a^2}\frac{\partial F}{\partial a},
\ee
and the heat capacity at constant volume
\be
\label{heatcapacity}
C_v= \frac{\partial E}{\partial T} = -T\frac{\partial^2 F}{\partial T^2}=-2\pi a x\partial_x^2 F.
\ee

We see that the classical term and the renormalization term 
in the free energy (\ref{74cc})
do not depend on the
temperature; therefore, the entropy and the heat capacity do not depend on those terms.
Therefore, the entropy and the heat capacity at constant volume are
given by the derivatives of the function $\Phi$,
\bea
S &=&\frac{\pi}{8}\partial_x \Phi,
\\
C_v &=& \frac{\pi}{8}x\partial_x^2\Phi.
\eea
By changing the integration variable $t\mapsto a^2t$ we rewrite the function $\Phi$ as
\bea
\Phi &=&\int\limits_0^\infty \frac{dt}{t^3}e^{-tz^2a^2}
\left\{\Omega\left(x^2t\right)W\left(t\right)
-R_{GR}\left(t\right)\right\}.
\label{711cxx}
\eea 
Differentiating the function $\Phi$ with respect to $x$, we get
\bea
\partial_x \Phi&=& 2x \int\limits_0^\infty \frac{dt}{t^2}e^{-ta^2z^2}\Omega'\left(x^2t\right)W(t)\,,
\\
\partial_x^2\Phi&=& 2\int\limits_0^\infty \frac{dt}{t^2}e^{-ta^2z^2}
\left\{\Omega'\left(x^2t\right)+2x^2 t\Omega''\left(x^2t\right)\right\}W(t).
\label{712cxx}
\eea

We will need the asymptotics of the function $\Omega(t)$
obtained in \cite{avramidi12a}. We have as $t\to 0$
\bea
\Omega(t)&=&1+2\exp\left(-\frac{\pi^2}{t}\right)
+O\left(e^{-4\pi^2/t}\right)\,,
\eea
and as $t\to \infty$,
\bea
\Omega(t)&=&\frac{1}{\sqrt{\pi}}\left[
t^{1/2}+2t^{1/2}e^{-t}
+O\left(e^{-4t}\right)\right]\,.
\eea

Because of the asymptotic behavior of $\Omega'$ as $t \to 0$, the 
integrals converge at $t \to 0$. The function  $W(t)$ increases exponentially
at infinity with exponent $e^t$ or $e^{2\lambda t}$, the function $\Phi$ has a
singularity for $z^2 < \max\left\{ \frac{1}{a^2}, 2\Lambda \right\}$. We may 
then view the maximum of these parameters as analogous to
$\Lambda_{QCD}$, an infrared cutoff below which our analysis ceases to
describe this system.

When $z$ is taken to zero, the integrals (\ref{711cxx}) and (\ref{712cxx}) do not converge,
and so we do not examine the free energy or entropy past this point. 
However, as we will see later, because the asymptotic behavior of 
$\Omega'\left(t\right)+2t\Omega''\left(t\right)$ as $t \to \infty$ is proportional
to $e^{-t}$, the heat capacity may converge even in the limit $z \to 0$.
We decompose $W$ according to
\bea
W(t)=\frac{4}{\sqrt{\pi}}t^{3/2}\left(
e^{2\lambda t}-8e^t-14\right)+V(t)\,,
\eea
where the function $V$ is exponentially small as $t\to \infty$.
We may then split the integral for $\partial_x^2\Phi$ into four parts:
\be
\partial_x^2\Phi = I_1 + I_2+I_3 +I_4\,,
\ee
where
\bea
I_1&=& \frac{8}{\sqrt{\pi}}\int\limits_0^\infty \frac{dt}{t^{1/2}}e^{-t(a^2z^2-2\lambda)}
\left\{\Omega'\left(x^2t\right)+2x^2 t\Omega''\left(x^2t\right)\right\},
\\
I_2&=& -\frac{64}{\sqrt{\pi}}\int\limits_0^\infty \frac{dt}{t^{1/2}}e^{-t(a^2z^2-1)}
\left\{\Omega'\left(x^2t\right)+2x^2 t\Omega''\left(x^2t\right)\right\},
\\
I_3&=& -\frac{112}{\sqrt{\pi}}\int\limits_0^\infty \frac{dt}{t^{1/2}}e^{-ta^2z^2}
\left\{\Omega'\left(x^2t\right)+2x^2 t\Omega''\left(x^2t\right)\right\},
\\
I_4&=& 2\int\limits_0^\infty \frac{dt}{t^{2}}e^{-ta^2z^2}
\left\{\Omega'\left(x^2t\right)+2x^2 t\Omega''\left(x^2t\right)\right\}V(t).
\eea

Notice that the function in the integrand is exponentially small at infinity, namely,
\be
\Omega'(t) + 2t\Omega''(t) = \frac{2t^{1/2}}{\sqrt{\pi}}(2t-3)e^{-t} + O(e^{-4t})
\label{722cc}
\ee
Therefore, the integrals $I_3$ and $I_4$ converge for any $x$.
The integral $I_1$ converges only for $x^2>2\lambda-a^2z^2$ and the integral $I_2$ converges
for $x^2>1-a^2z^2$.
Allowing the infrared cutoff $z$ to go to zero, 
the integral $I_1$ converges only for $x^2>2\lambda$ and the integral $I_2$ converges
for $x^2>1$. 

Therefore, all of the integrals converge at high temperature but the heat capacity
has a singularity either at the temperature $x_c=\sqrt{2\lambda}$ (for positive $\lambda>1/2$) or
at $x_c=1$ if $\lambda<1/2$ (including the case of negative 
cosmological constant $\Lambda$). Recalling that $\lambda=a^2 \Lambda$ and
$x=2\pi a T$, this defines the critical temperature
\be
T_{c}=\sqrt{\max\left\{\frac{\Lambda}{2\pi^2},\frac{1}{4\pi^2 a^2}\right\}},
\label{723vv}
\ee
below which the system will undergo a phase transition.
Notice that the smallest value of the critical temperature is
\be
T_{c, {\rm min}}=\sqrt{\max{\left\{\frac{\Lambda}{2\pi^2},0\right\}}}.
\label{723bb}
\ee
The phase diagram 
 of the graviton gas for the positive cosmological constant
has the form illustrated on the graph
Figure 1.

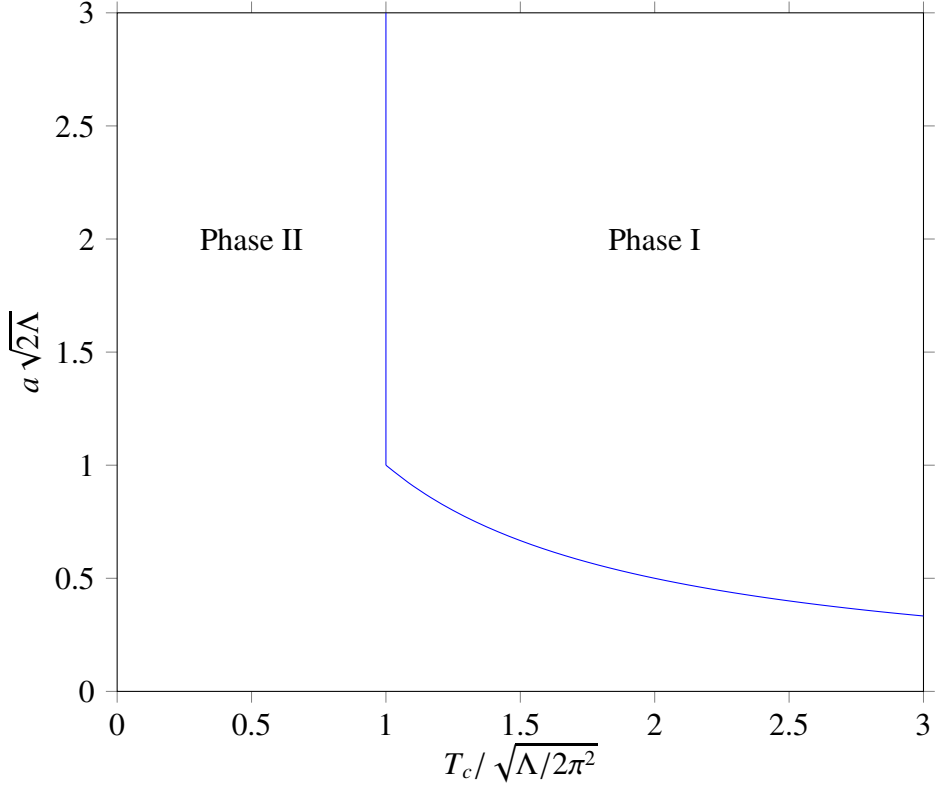
\begin{figure}[h!]
\begin{tikzpicture}
	\begin{axis}[width=350pt, tick align=outside, xmin = 0, ymin = 0, xmax = 3, ymax = 3,
	             xlabel = {$T_c/\sqrt{\Lambda / 2\pi^2}$}, ylabel = {$a\sqrt{2\Lambda}$}]
		\addplot+[name path = radius,mark=none,smooth,domain = 1:3, blue] {1/x};
		\addplot+[name path = Lambda,mark=none,smooth, blue] coordinates {(1,1) (1,3)};
		\node at (axis cs:0.5,2) {Phase II};
		\node at (axis cs:2,2) {Phase I};	
	\end{axis}
\end{tikzpicture}
\caption{Phase diagram of the graviton gas}
\end{figure}

We also study the high temperature limit as $x\to \infty$.
The asymptotic behavior of the combination of derivatives of $\Omega$ (\ref{722cc}) implies 
that the high temperature limit corresponds to the limit of 
$t \to \infty$. We find the limit of $I_1$ through $I_3$ replacing the $\Omega$ functions 
by their leading asymptotics and integrating:
\bea
I_1 &\sim& -\frac{16}{\pi x},
\\
I_2 &\sim& \frac{128}{\pi x},
\\
I_3 &\sim& \frac{224}{\pi x}.
\eea
The integral $I_4$ is evaluated by changing variables $t \to t/x^2$ and using
$V(0) = W(0) = 2$; we get
\be
I_4 \sim 4x^2 \nu\,,
\ee
where $\nu$ is the constant defined by the integral
\be
\nu = \int\limits_0^\infty \frac{dt}{t^2}
\left\{\Omega'\left(t\right)+2 t\Omega''\left(t\right)\right\}.
\ee
The integral $I_4$ dominates in the high temperature limit and determines the heat capacity per volume
\be
\frac{C_v}{V} \sim \frac{\nu_1}{4\pi a^3}x^3 = 2\pi^2\nu T^3 \,.
\ee
The $T^3$ dependence is characteristic of the photon gas and,
as has been found in our previous paper
\cite{avramidi12b},
of the gluon gas as well.

Next we study the behavior of the heat capacity near the critical
temperature.
By using eq. (\ref{722cc}) and setting $z=0$ we get
\bea
I_1 &\sim& \frac{32x^3}{\pi}\int_0^\infty dt\;t e^{-t(x^2-2\lambda)},
\\
I_2 &\sim& -\frac{256x^3}{\pi}\int_0^\infty dt\;t e^{-t(x^2-1)}\,.
\eea
We obtain, as $x \to \sqrt{2\lambda}^+$,
\bea
I_1 &\sim&  \frac{32\lambda}{\pi} \left(x - \sqrt{2\lambda}\right)^{-1},
\eea
and as $x \to 1^+$,
\bea
I_2 &\sim&  -\frac{128}{\pi} (x - 1)^{-1}.
\eea
The critical exponent of $(-1)$ is indicative of a second-order phase transition. 

The temperature at which the phase transition occurs depends on the value of 
the cosmological constant.
In the case that $\Lambda>(2a^2)^{-1}$, the phase transition occurs at 
the temperature $T_c=\sqrt{\Lambda/(2\pi^2)}$
and if $\Lambda \le (2a^2)^{-1}$ (also if $\Lambda<0$)
the phase transition occurs at the temperature $T_c=1/(2\pi a)$.
The asymptotics of the heat capacity near the critical temperature are:
if $\Lambda>(2a^2)^{-1}$ then as $T\to T_c$
\be
C_v \sim \frac{2^{3/2}}{\pi}\, a^2\Lambda^{3/2}\left(T-T_c\right)^{-1} \,,
\ee
with $T_c=\sqrt{\Lambda/(2\pi^2)}$.
Since the heat capacity at constant volume as a function of temperature grows at infinity,
this means that the heat capacity must have a minimum at some
temperature $T_1>T_c$.
Further, if $\Lambda<(2a^2)^{-1}$ then as $T\to T_c$
\be
C_v \sim -\frac{8}{\pi a}(T-T_c)^{-1} \,,
\ee
with $T_c=1/(2\pi a)$;
in the case that $\Lambda=(2a^2)^{-1}$ we have
\be
C_v \sim -\frac{7}{\pi a}(T-T_c)^{-1},
\ee
with $T_c=1/(2\pi a)$. This means that the heat capacity must vanish at some
temperature $T_2>T_c$ (see Figure 2.).

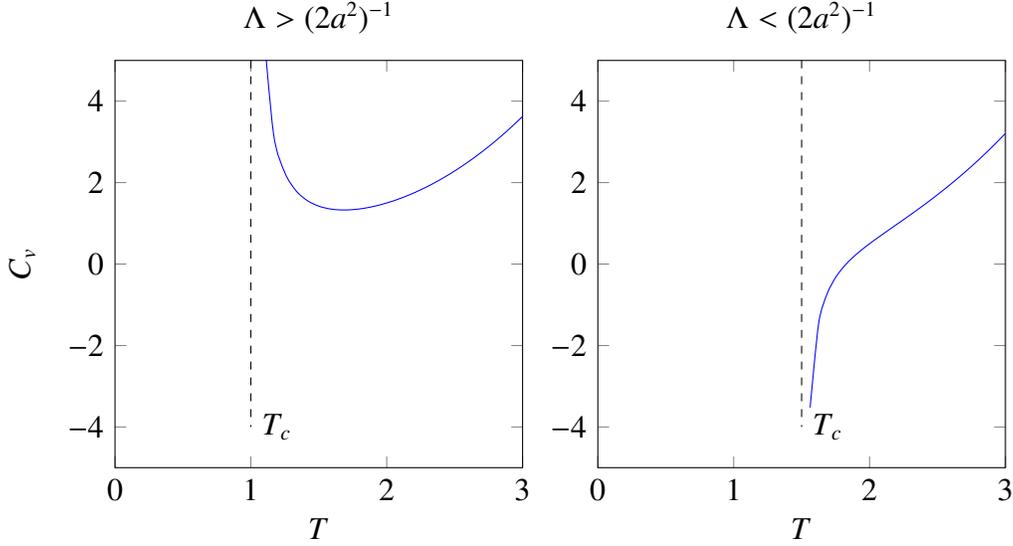
\begin{figure}[h!]
\begin{tikzpicture}
    \begin{groupplot}[group style={group size=2 by 2},height=7cm,width=7cm]

    \nextgroupplot[%
     xmin = 0, xmax = 3, ymin=-5, ymax=5, title = {$\Lambda>(2a^2)^{-1}$},
xlabel = {$T$}, ylabel = {$C_v$}
]
    \addplot+[mark=none,smooth,domain = 1:3, blue] {0.5/(x-1)+x^3/8};
    \addplot+[mark=none, dashed, black] coordinates {(1,5) (1,-4)} node [right] {$T_c$};

    \nextgroupplot[  xmin = 0, xmax = 3, ymin=-5, ymax=5, title = {$\Lambda<(2a^2)^{-1}$},
xlabel = {$T$}]
    \addplot+[mark=none,smooth,domain = 1.5:3, blue] {-0.25/(x-1.5)+x^3/8};
    \addplot+[mark=none, dashed, black] coordinates {(1.5,5) (1.5,-4)} node [right] {$T_c$};

    \end{groupplot}
\end{tikzpicture}
\caption{Heat capacity as a function of temperature}
\end{figure}


\section{Discussion}
\setcounter{equation}{0}

It is well-known that the gravitational action is unbounded from below and unstable.
The primary goal of this paper was to study the quantum gravitational field restricted
to a set of manifolds which have an action that is classically bounded from
below, and then examine how one-loop quantum effects disturb that stability.
In order to calculate the one-loop effective action exactly, it is
necessary to study a spacetime with a great degree of symmetry.
We studied the thermal Einstein universe $S^1 \times S^3$ with non-zero 
cosmological constant, varying the model only with respect to the radii, $a_1$ and $a$,
of the circle $S^1$ and the sphere $S^3$ respectively.
This spacetime is off-shell, so, strictly speaking,
it does depend on the gauge of the quantum field. However, we used the generally
accepted minimal covariant De Witt's gauge in which all operators become
Laplace type.

We computed the exact trace of the heat kernels of all relevant operators,
which enabled us to calculate the one-loop effective action exactly. The lowest
value of the of the graviton operator can be chosen to be positive by 
adjusting the cosmological constant, but the ghost operator always yields a
negative eigenvalue, indicating an unstable mode for any radius of the 
Einstein universe. This may indicate a problem with the gauge condition and requires
a detailed further study.

We also studied the thermal properties of the model. We found
that while the free energy and entropy are ultraviolet divergent, the heat
capacity is well-defined even in the infrared limit. In the high-temperature
limit, the heat capacity of the graviton gas has a $T^3$ dependence which is 
typical of a photon gas, and has also been found in our previous paper 
\cite{avramidi12b}
to be
consistent with a gluon gas. 

We also computed the asymptotics of the heat capacity near the critical temperature
and found that the heat capacity has a branching
singularity $\sim (T-T_c)^{-1}$ at a finite critical temperature 
$T_{c}$ given by (\ref{723vv}).

In the case of negative or small positive cosmological constant,
$\Lambda<(2a^2)^{-1}$, the system exhibits a rather anomalous peculiar behavior
with the negative heat capacity due to the presence of the unstable mode of the
ghost operator.
It is common in bound gravitational systems to have negative heat capacity. 
For instance, the temperature of a black hole decreases as heat is added to it.
The fact that the heat capacity changes sign at 
some temperature $T_2$ indicates that the system has a minimum internal
energy at that temperature. 

It is interesting to play with the minimal value of the critical temperature
given by (\ref{723bb}).
If we substitute the observed value of the cosmological constant, 
$\Lambda \sim 10^{-52} m^{-2}$, then the minimum critical temperature
is approximately 
\be
T_{c, {\rm min}}=\frac{\hbar c }{k_B}\sqrt{\frac{\Lambda}{2\pi^2}} \sim 5\times 10^{-4} K.
\ee
One can speculate that if the universe
cools below the critical temperature, it is likely that some 
degrees of freedom would be frozen leaving a cosmic background thermal
graviton radiation with temperature $T_c$.

The techniques used in this paper are very general. It would be interesting
to extend this model to higher-derivative quantum gravity or to supergravity,
in which the one-loop action vanishes on-shell.


\end{document}